\documentclass[12pt]{article}

\usepackage{amstext,amsmath,amssymb,amsfonts,bbm,slashed}
\usepackage[latin1]{inputenc}
\usepackage{epsfig}
\usepackage{hyperref}
\usepackage{amsthm}
\usepackage{subfigure}
\usepackage{multirow}

\newcommand{\kin}{\text{kin\,}}
\newcommand{\ext}{\text{ext\,}}
\newcommand{\inter}{\text{int\,}}

\def\C{{\mathbbm C}}
\def\N{{\mathbbm N}}

\def\R{{\mathbbm R}}

\newcommand{\cG}{{\mathcal G}}
\newcommand{\bG}{{\partial\mathcal G}}
\newcommand{\tJ}{{\widetilde{J}}}

\newcommand{\cexG}{\mathcal G_{\text{color}}}
\newcommand{\bJ}{ J_{\partial} }

\newcommand{\cL}{{\mathcal L}}
\newcommand{\cV}{{\mathcal V}}
\newcommand{\cF}{{\mathcal F}}

\newcommand{\um}{{\underline m}}

\newtheorem{lemma}{Lemma}

\newtheorem{definition}{Definition}
\newtheorem{theorem}{Theorem}

\newcommand{\bea}{\begin{eqnarray}}
\newcommand{\eea}{\end{eqnarray}}
\newcommand{\beq}{\begin{equation}}
\newcommand{\eeq}{\end{equation}}

\makeatletter

\begin{document}

\begin{titlepage}
\begin{flushright}
pi-qg-245\\
Lpt-Orsay-11-99\\
ICMPA-MPA/2011/017
\end{flushright}

\vspace{20pt}

\begin{center}

{\Large\bf A Renormalizable 4-Dimensional \\
\medskip
Tensor Field Theory}
\vspace{15pt}

{\large Joseph Ben Geloun$^{a,c,\dag}$ and Vincent Rivasseau$^{b,\ddag} $}

\vspace{15pt}

$^{a}${\sl Perimeter Institute for Theoretical Physics}\\
{\sl 31 Caroline St. N., ON, N2L 2Y5, Waterloo, Canada}\\
\vspace{5pt}

$^{b}${\sl Laboratoire de Physique Th\'eorique, CNRS UMR 8627}\\
{\sl Universit\'e Paris-Sud, 91405 Orsay, France}
\vspace{5pt}

$^{c}${\sl International Chair in Mathematical Physics and Applications\\ (ICMPA-UNESCO Chair), University of Abomey-Calavi,\\
072B.P.50, Cotonou, Rep. of Benin}\\
\vspace{5pt}
E-mails:  {\sl $^{\dag}$jbengeloun@perimeterinstitute.ca, 
$^\ddag$rivass@th.u-psud.fr}

\vspace{10pt}

\begin{abstract}
We prove that an integrated version of the Gurau colored tensor model supplemented with the usual Bosonic propagator on $U(1)^4$
is renormalizable to all orders in perturbation theory. The model is of the type
expected for quantization of space-time in $4D$ Euclidean gravity and
is the first example of a renormalizable model of this kind. Its vertex and 
propagator are four-stranded like in $4D$ group field theories, but without gauge averaging on the strands.
Surprisingly perhaps, the model is of the $\phi^6$ rather than of the $\phi^4$ type, since two different $\phi^6$-type interactions are 
log-divergent, i.e. marginal in the renormalization group sense.
The renormalization proof relies on a multiscale analysis. It identifies all divergent graphs through a power counting theorem. 
These divergent graphs have internal and external structure of a particular kind called melonic.
Melonic graphs dominate the 1/N expansion of colored tensor models and generalize the planar ribbon graphs of matrix models. 
A new locality principle is established for this category of graphs which allows to renormalize their divergences 
through counterterms of the form of the bare Lagrangian interactions.
The model also has an unexpected anomalous log-divergent $(\int \phi^2)^2$ term, which can be interpreted as
the generation of a scalar matter field out of pure gravity.
\end{abstract}
\end{center}

\noindent  Pacs numbers:  11.10.Gh, 04.60.-m
\\

\setcounter{footnote}{0}

\end{titlepage}

\section{Introduction}

The standard model is built out of renormalizable 4 dimensional quantum field theories. In the Wilsonian point of view
this is natural since these theories have long-lived logarithmic
flows. They can survive almost unchanged through long sequences of renormalization group transformations.
Our universe seems to favor such theories because it is very large 
(at least in terms of the Planck scale). It would be desirable to describe also quantum gravity with a similar renormalizable model  \cite{Rivasseau:2011xg}.

However the ordinary approach to quantize the Einstein-Hilbert action around flat space
is well known to lead to a perturbatively non-renormalizable theory. Attention has turned to add symmetries and 
extended objects (supergravity, superstring and related approaches) or to create space-time itself from more fundamental entities.
In this second point of view these entities can still obey the rules of a more abstract ``pregeometric" quantum field theory but it is natural
to drop some of the fundamental axioms, for instance ordinary locality and Poincar\'e invariance. Let us from now on restrict ourselves to Euclidean quantum field theory.
The most natural assumption is that classical Euclidean space-time and general relativity could be an effective product of such a pregeometric quantum field theory 
resulting from a phase transition, just as hadronic physics is the effective product of QCD.

Until now the main success in this direction is the random matrix approach to the quantization of 2$D$ gravity \cite{Di Francesco:1993nw}.
It produces indeed a theory of continuous Riemann surfaces through a phase transition with computable critical exponents
\cite{mm,Kaz}. It can also reproduce through a double scaling
a sum of surfaces of different genera \cite{BKaz,DS,GM}.
  Moreover the theory allows fruitful applications to 2$D$ statistical physics through the KPZ map
\cite{KPZ,D1,DK,Dup}. This success relies on a fundamental tool to analyze the statistical properties of large random matrices, namely the $1/N$ expansion \cite{Hooft}. 

Random tensor models \cite{ambj3dqg, mmgravity, sasa1, Ambjorn:1991wq} of rank $D\ge 3$ are the first and most natural attempt to generalize this success to higher dimensions $D\ge 3$.
The natural vertex is the $D$-stranded $\phi^{D+1}$-type vertex associated to the complete graph on $D+1$ points. Since such points define a $D$-simplex, 
the Feynman graphs of the theory are dual to triangulations of $D$-dimensional topological spaces.
However until recently there was no way to address statistical properties of large tensors of rank higher than 2 through a $1/N$ expansion, hence such 
models have been mostly studied through computer simulations.
See also \cite{Sasakura:2011sq,Sasakura:2011nj,Rey:2010uz} for other related approaches.

Group field theory is a special kind of random tensor model in which one adds 
a Lie group $G$ and a gauge invariance \cite{Boul,Ooguri:1992eb,laurentgft,quantugeom2,Oriti:2011jm}\footnote{Group field theory is related to loop quantum gravity since the Feynman amplitudes
of group field theory are the spinfoams in the covariant version of LQG. But it improves the latter with key
ingredients: canonical combinatoric weights for the spinfoams from Wick theorem, plus the potential to harness the power of quantum field theory tools:
functional integrals, non-perturbative expansions and the renormalization group.}.
In dimension $D$ the natural Lie group is $SO(D)$ (or its covering group).  Gauge invariance consists in averaging over a single simultaneous action of $G$ 
on all $D$-strands of the propagator\footnote{In the initial paper of Boulatov and many subsequent works, the propagator is
incorporated into the vertex, an unfortunate convention from the QFT point of view.}. This gauge invariance implements the flatness condition of the $BF$ theory, 
because it ensures trivial holonomy for parallel transport of vectors along all faces of the triangulated space. In three dimensions it seems
related to the quantization of gravity because the classical Einstein-Hilbert action reduces to the $BF$ theory in $D=3$.

In four dimensions more elaborate propagators have been proposed \cite{BC1,Engle:2007qf,Freidel:2007py,Geloun:2010vj}
to implement the Plebanski simplicity constraints on the 4 dimensional Ooguri GFT or $4D$ $BF$ theory. Hopefully, 
this could free the local modes of classical $4D$ gravity,
those responsible for gravitational waves. However, the analysis of the
corresponding amplitudes has turned out to be harder than expected
\cite{Kraj}. Detailed studies for the power counting of group field theory amplitudes \cite{FreiGurOriti, sefu1, sefu2, sefu3, BS1, BS2, BS3} 
have not lead to any renormalizable group field theory yet.

Recently a breakthrough occurred. A new
class of colored models \cite{color} provided at last tensor theories and group field theories 
with their missing analytic tool, namely the $1/N$ expansion \cite{Gur3,GurRiv,Gur4}. 
Results on statistical mechanics
on random $D  \ge 3$ geometries followed quickly  \cite{Bonzom:2011zz, Bonzom:2011ev, Benedetti:2011nn}.
We refer to \cite{Gurau:2011xp} for a review and to \cite{Gurau:2009tz, Gurau:2010nd, Baratin:2011tg, Gurau:2011sk} for other results or aspects of  this thriving subject.

Even more important perhaps,
actions for \emph{uncolored} random tensor theories were developed \cite{Gurau:2011tj}. 
They obey an infinite dimensional symmetry algebra, based on $D$-ary trees and their fusion rules, which we propose to call the Gurau algebra.
This theory and this symmetry 
has been proved universal in the precise sense of probability theory: every independent-identically-distributed or even \emph{invariant} probability law on
\emph{uncolored}  random tensors
is governed in the large $N$ limit by the $1/N$ expansion of colored models \cite{Gurau:2011kk}. Hence it is the correct extension
in higher dimensions of the central limit theorem and of the Wigner-Dyson theory of random matrices.

This breakthrough opens a new program, namely the systematic investigation of 
tensor field theories of rank higher than 2 and the classification of their renormalization group flows and critical points using these new analytic tools. 
We hope this could lead to a simpler and more convincing quantization of gravity in 3 and 4 dimensions. It could also provide 
the correct extension to higher dimensions of $2D$ conformal symmetry and integrability which have been
so useful in the study of $2D$ statistical mechanics models and of their phase transitions. 

This paper is a first step in this  program. We use the  $1/N$  expansion of colored random tensors to build 
the first renormalizable \emph{uncolored} rank 4 tensor quantum field theory. Our model can be considered as a natural higher rank analog of the
Grosse-Wulkenhaar model  \cite{Grosse:2004yu,Rivasseau:2005bh}, which was built around the ordinary  $1/N$  expansion of random matrices. 
It can be also considered as a group field theory with group
$G = U(1)$  but we prefer not to use this terminology since we perform  no gauge averaging on the propagator strands.
Our model is a four dimensional quantum field theory of a single scalar field with the ordinary $(- \Delta + m^2)^{-1}$ propagator. 
For earlier approaches to group field theory with inverse Laplacian propagators see \cite{Oriti:2010hg} and references therein. 
We also mention that the requirement of Laplacian dynamics 
in a renormalization analysis of group field theory has been 
underlined in \cite{Geloun:2011cy}.
Each coordinate is associated to a tensor index, so the model is \emph{both 4 dimensional and rank 4}.
 For simplicity, we choose to formulate the theory on the four dimensional torus  
$T_4=U(1)^4$ rather than on $\R^4$, but this is not a fundamental feature \footnote{
The compact four dimensional space $T_4=U(1)^4$ on which the theory lives could be replaced by $\R^4$. This would introduce the usual distinction between infrared and ultraviolet
divergences. The infrared divergences could be cured by an infrared regulator such as an harmonic potential \`a la Grosse-Wulkenhaar.
This is left to a future study.}.

Only the interaction of our model is new.
We obtain this interaction by truncating 
the infinite series of melonic terms in the Gurau action \cite{Gurau:2011tj} to eliminate the irrelevant terms.
This parallels exactly what is done on the infinite series of local interaction terms $\int \phi (x)^n dx$
in ordinary renormalizable quantum field theory. The usual  $\int \phi^4(x)dx$ action is the correct truncation
for renormalizability  in 4 dimensions. Irrelevant terms do not appear in the bare action.
It is also the recipe for renormalizable matrix-like quantum field theories such as the Grosse-Wulkenhaar model, where
the infinite series of Tr$M^n$ terms is truncated (in that model again to order 4).
This is because the Tr$ M^n$ terms are the right matrix analogs of local interactions. Similarly
melonic terms are the right analogs of local interactions for tensor theories of rank 3 or more\footnote{We nevertheless agree that when written in terms of coordinates on $T^4$
the melonic interactions look unfamiliar at first sight. The four different coordinates of $U(1)^4$ correspond to different strands in the propagator 
which are identified according to the melonic drawings. The resulting interaction is certainly not local in the usual sense.}. 

It happens that 6 is the right order of truncation for this model  
to get just renormalizability with the
$(- \Delta + m^2)^{-1}$ propagator. We need also to add the
correct terms of order 4 and 2. The theory  
generates a single unexpected  $(\int \phi (x)^2 dx )^2$  anomaly
which could be interpreted as the generation of a scalar matter field out of pure gravity (see Subsection 6.3). Adding the corresponding fourth order term to the Lagrangian
we prove through a multiscale analysis that our model is renormalizable to all orders of perturbation theory.

Our model being defined on a compact space there is only one half-direction for the renormalization
group. According to the usual quantum field theory conventions
we call it the ultraviolet direction, as it describes short range fluctuations of the field 
\footnote{Recall that the large Fourier modes in group field theory or  spin-foams
are often considered the infrared direction, because of a different interpretation.
If space-time is the effective product of a phase transition, this interpretation may be dubious.}.

Section 2 introduces the model and notations and states our main theorem. Section 3 writes its multiscale decomposition and bounds.
Sections 4 and 5 identify the contributions to renormalize, including the anomalous term. 
Section 6 performs renormalization through suitable Taylor expansions around the local melonic parts of 
every divergent subgraph in the multiscale analysis. Section 7 lists some perspectives and open problems.
An appendix  provides some details on calculations invoked in
the text and introduces a similar just renormalizable theory in dimension 3. The important physical issues of the underlying model symmetries, renormalization group flow 
and possible phase transitions of such models are postponed to subsequent works.

\section{The Model}
\label{sect:model}

 We start by a blitz review of the basic ingredient, colored rank 4 tensor field theory \cite{color}.  
Let us consider a family of $5=4+1$ complex fourth rank tensor fields  over the group
$U(1)$, $\varphi^a : U(1)^4 \to \mathbb{C}$. They are labeled with an index 
$a=0,1,2,\dots,4$, called color. These colored fields can be
expanded into Fourier modes 
\bea
\varphi^a_{1,2,3,4}
= \sum_{p_j \in \mathbb{Z}} \varphi^a_{[p_j]} e^{ip_1 \theta_1} e^{ip_2 \theta_2}
 e^{ip_3 \theta_3} e^{ip_4 \theta_4}\;,
\qquad \theta_i \in [0,2\pi)\;,\qquad
[p_j]= (p_1,p_2,p_3,p_4)\;.
\eea
where the group elements $h_i \in U(1)$. We adopt the notation $\varphi^a(h_1,h_2,h_3,h_4)=$
$ \varphi^a_{1,2,3,4}$. 
Remark that no symmetry under permutation of arguments 
is assumed for any of the field $\varphi^a$ 
and for the corresponding tensors $\varphi^a_{[p_j]}$. 

The kinetic part of the action for the last four fields is the 
standard ``local" colored one
\bea
S^{\kin, 1,2,3,4}=
 \sum_{a=1}^4 \int_{h_{j}}
\bar\varphi^a_{1,2,3,4}\varphi^a_{1,2,3,4}\; .
\eea
The symbol $\int_{h_j}$ stands for the
Haar measure over all group variables with label of the form $h_{j}$.
For each variable, this is merely the normalized compact integral $(1/2\pi)\int_{0}^{2\pi} d\theta_j $. 

The interaction part of the action is the standard colored action in 4 dimensions  \cite{color}
\bea
\label{eq:interaction}
S^{\inter}&=& \tilde\lambda
\int_{h_{ij}} 
\varphi^0_{1,2,3,4} \,
\varphi^1_{4,5,6,7} \,
\varphi^2_{7,3,8,9} \,
\varphi^3_{9,6,2,10}\,
\varphi^4_{10,8,5,1}
\crcr
&+&\bar{\tilde{\lambda}}
\int_{h_{ij}} 
\bar\varphi^0_{1,2,3,4}\,
\bar\varphi^1_{4,5,6,7} \,
\bar\varphi^2_{7,3,8,9}\,
\bar\varphi^3_{9,6,2,10}\,
\bar\varphi^4_{10,8,5,1}\; ,
\label{interacol}
\eea
where $\tilde \lambda$ and $\tilde{\bar\lambda}$
are coupling constants.

We want to build a model in which the field with color $0$
is singled out and is the only dynamical field. Hence for that single field we introduce a different propagator
with quadratic action
\bea
S^{\kin,0} =
 \int_{h_{j}}
\bar\varphi^0_{1,2,3,4}
\Big(-\sum_{s=1}^4 \Delta_{s} + m^2\Big)\varphi^0_{1,2,3,4} \;,
\eea
where $\Delta_{s}:=\partial^2_{(s)\,\theta}$ 
denotes the Laplacian on $U(1)\equiv S^1$ acting on 
the strand index $s$. The corresponding Gaussian measure 
of covariance $C = (-\sum_s \Delta_s + m^2)^{-1}$ is noted as $d\mu_C$.

We integrate over the four colors 1,2,3,4  and obtain a partition function with an
effective action for the last tensor $\varphi^0$ \cite{Gurau:2011tj}: 
\bea
&&
Z = \int d\mu_C[\varphi^0]\; e^{- S^{\inter,0}  }\;,  \crcr
&&
S^{\inter, 0} = 
\sum_{\mathcal B} 
\frac{(\tilde\lambda\bar{\tilde{\lambda}})_{\mathcal B} }{
\text{Sym}(\mathcal B)} N^{f(p,D) -\frac{2}{(D-2)!}\omega(\mathcal B)}
\text{Tr}_{\mathcal B} [\bar\varphi^0\varphi^0]\;,
\label{integrateco}
\eea
where the sum in $\mathcal B$ is performed on all bubbles,
or connected vacuum graphs with colors $1$ up to $D$
and $p$ vertices; $f(p,D)$ is a positive function of the number of
vertices and the dimension; $\omega(\mathcal B) := \sum_{J} g_{J}$
is the sum of genera of sub-ribbon graphs called jackets $J$
of the bubble, and
$\text{Tr}_{\mathcal B}[\bar\varphi^0\varphi^0]$ are called
tensor network operators.
Graphs with $\omega(\mathcal B) =0$ are called
melons. Non melonic contributions defined
by $\omega(\mathcal B) >0$ are clearly suppressed from (\ref{integrateco}).
For details on all this terminology we refer to \cite{Gurau:2011xp} and references therein.  
We will concentrate only on the melonic sector of the theory. A fundamental idea of  \cite{Gurau:2011tj}
is to attribute a different coupling constant to different  tensor network operators. 
We simply write (dropping from now on the last color index $0$)
\bea
S^{\inter, 0} = 
\sum_{\mathcal B} 
\frac{\lambda_{\mathcal B}}{
\text{Sym}(\mathcal B)} 
\text{Tr}_{\mathcal B} [\bar\varphi\varphi]\;.
\eea
In order to get a renormalizable theory,
we have to truncate this action to a finite number of 
marginal and relevant terms, in renormalization group language.

The trace operators or effective interaction terms that we 
will consider in the following are monomials of order six at most, 
given by
\bea \label{S62}
S_{6;1} &=& \int_{h_{j}}
\varphi_{1,2,3,4} \,\bar\varphi_{1',2,3,4}\,\varphi_{1',2',3',4'} \,\bar\varphi_{1'',2',3',4'} \,
\varphi_{1'',2'',3'',4''} \,\bar\varphi_{1,2'',3'',4''} \crcr
&&+ \text{permutations } \;,
\\
S_{6;2} &=& \int_{h_{j}}
\varphi_{1,2,3,4} \,\bar\varphi_{1',2',3',4}\,\varphi_{1',2',3',4'} \,\bar\varphi_{1'',2,3,4'}\,
\varphi_{1'',2'',3'',4''} \,\bar\varphi_{1,2'',3'',4''}\, \cr\cr
&&+ \text{permutations }\;,
\label{interac6} 
\\    \label{S41}
S_{4;1} & =&  \int_{h_j} \varphi_{1,2,3,4} \,\bar\varphi_{1',2,3,4}\,\varphi_{1',2',3',4'} \,\bar\varphi_{1,2',3',4'}\, 
+ \text{permutations }  \;,
\eea
where the sum is over all 24 permutations of the four color indices. 

\begin{figure}
 \centering
     \begin{minipage}[t]{.8\textwidth}
      \centering
\includegraphics[angle=0, width=5cm, height=1cm]{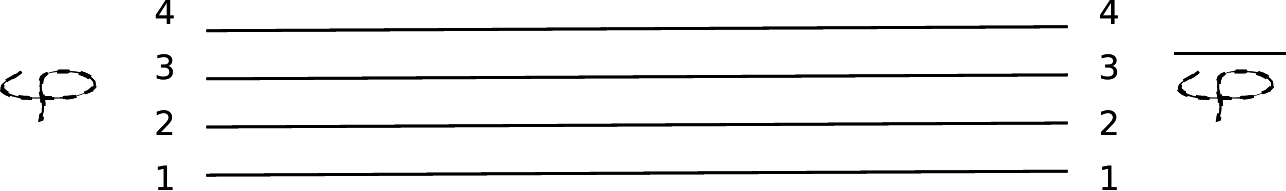}
\caption{ {\small The propagator. }}
\end{minipage}
\end{figure}

\begin{figure}
 \centering
     \begin{minipage}[t]{.8\textwidth}
      \centering
\includegraphics[angle=0, width=11cm, height=4cm]{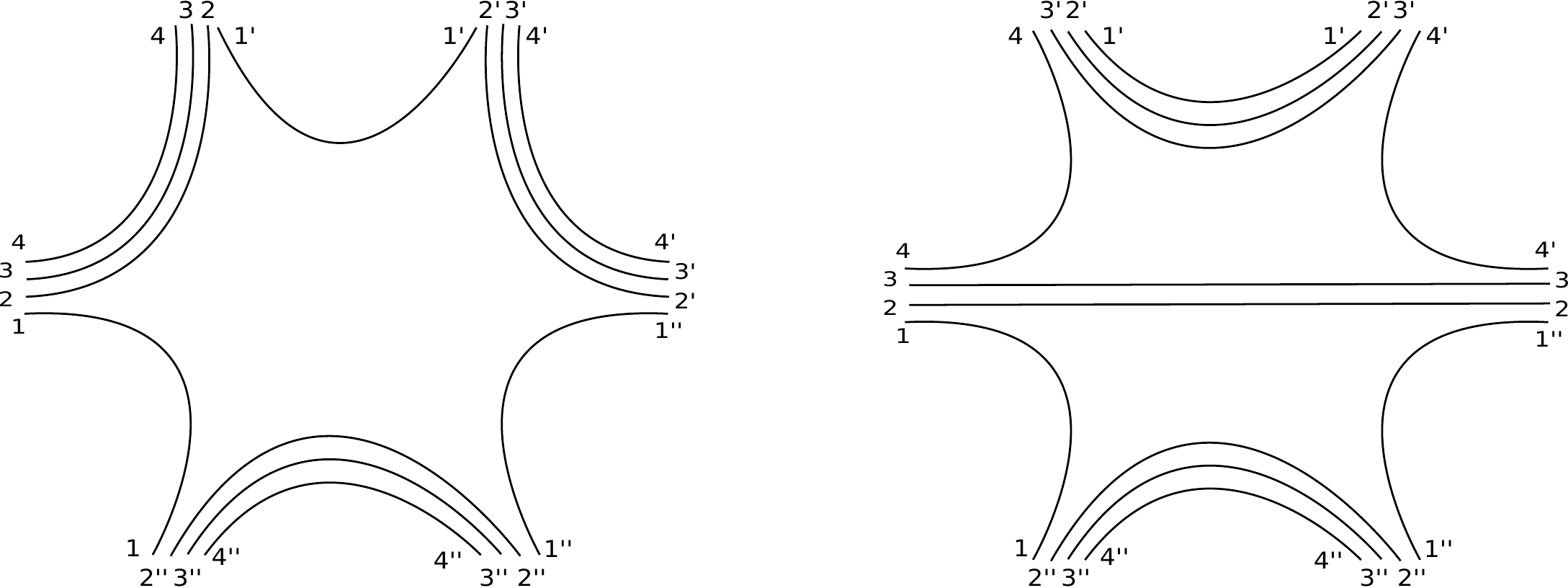}
\caption{ {\small Vertices of the type $V_{6;1}$ (left)  and $V_{6;2}$ (right). \label{vert6}}}
\end{minipage}
\end{figure}

\begin{figure}
 \centering
     \begin{minipage}[t]{.8\textwidth}
      \centering
\includegraphics[angle=0, width=10cm, height=5cm]{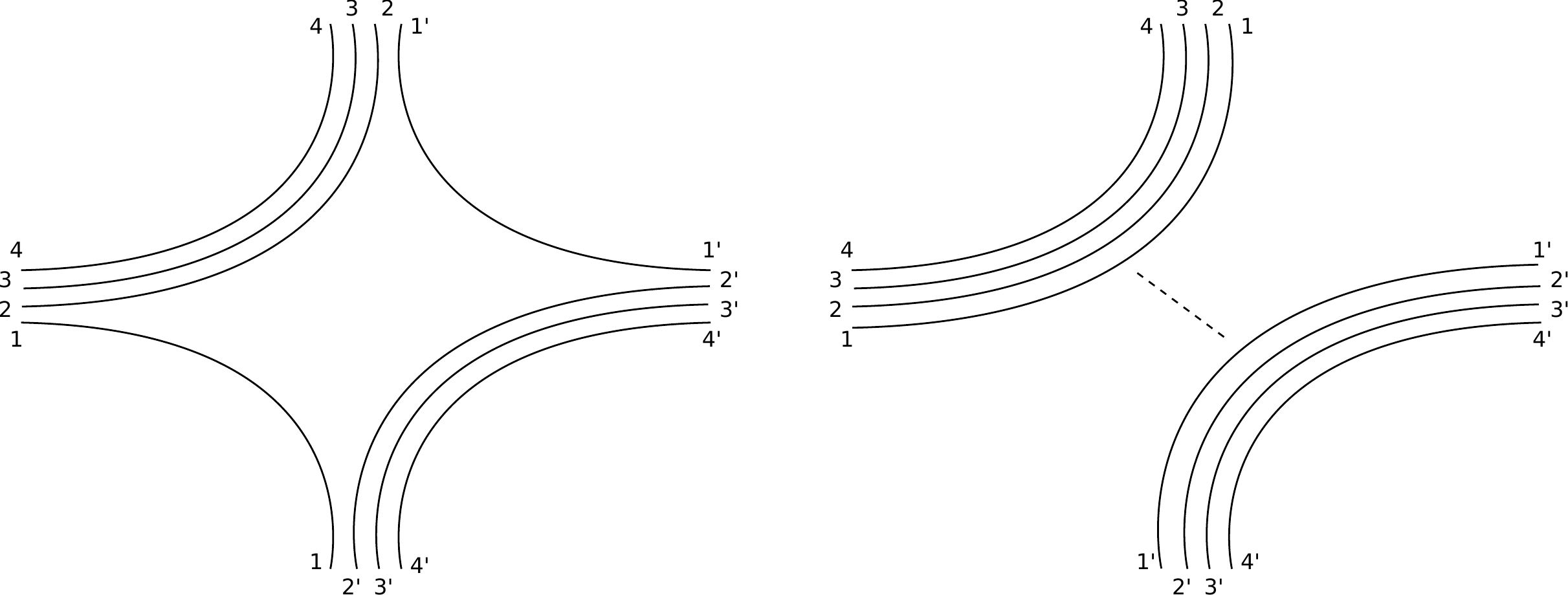}
\caption{ {\small Vertices of the type $V_{4;1}$ (left)  and $V_{4;2}$ (right). 
\label{vert4}}}
\end{minipage}
\end{figure}

Feynman graphs are tensor like: fields are represented by 
half lines with four strands, propagators are lines 
with the same structure (see Fig.1), meanwhile, vertices 
are non local objects as depicted in Fig.\ref{vert6} and \ref{vert4}. 
For convenience, we will sometimes use simplified diagrammatics 
where the strand structure will be hidden. 

Furthermore, the renormalization analysis of Section \ref{sect:renorm}
also leads to add to the action another $\varphi^4$-type anomalous divergent term, namely:
\bea
S_{4;2}&=&
\left[\int_{h_j} \bar\varphi_{1,2,3,4} \,
\varphi_{1,2,3,4} \right]
\left[\int_{h'_j} \bar\varphi_{1',2',3',4'}\, 
\varphi_{1',2',3',4'} \right]   .
\label{fifi}
\eea
The interaction (\ref{fifi}) can be considered as a joined pair of two 
factorized $\varphi^2$ vertices which we represent as two lines 
with a dotted line between them, see Fig.\ref{vert4}.

In the next section we shall introduce an ultraviolet cutoff $\Lambda$ on the propagator, which becomes
$C^{\Lambda}$. We introduce as usual bare and renormalized couplings, the difference of which are coupling constants 
counterterms, called $CT$.
We have also to introduce counterterms in the bare action 
to perform the mass and wave function renormalization hence we also define
quadratic terms in the action:
\beq
S_{2;1}= \int_{h_{j}}   \bar\varphi_{1,2,3,4}\varphi_{1,2,3,4}   \;,
\qquad 
S_{2;2}= \int_{h_{j}} \bar\varphi_{1,2,3,4}
\Big(-\sum_{s=1}^4 \Delta_{s}\Big)\varphi_{1,2,3,4} \;.
\label{quadrat}
\eeq
The propagator $C$ has for coefficients the renormalized mass $m^2$ and
the renormalized wave function $1$.

The action of the model is then defined as
\bea
S^{\Lambda} =  \lambda^{\Lambda}_{6;1}   S_{6;1} +   \lambda^{\Lambda}_{6;2}  S_{6;2} +
\lambda^{\Lambda}_{4;1}  S_{4;1} +  \lambda^{\Lambda}_{4;2} S_{4;2} +  CT^{\Lambda}_{2;1} S_{2;1} +  CT^{\Lambda}_{2;2} S_{2;2}
\eea
and the partition function is
\bea
Z = \int d\mu_{C^{\Lambda}}  [\varphi]\; e^{- S^{\Lambda}  } .  \label{baremodel}
\eea

The renormalization theorem means that we can define four renormalized coupling constants
$ \lambda^{\text{ren}}_{6;1}, \lambda^{\text{ren}}_{6;2}, \lambda^{\text{ren}}_{4;1}, \lambda^{\text{ren}}_{4;2}$
such that choosing appropriately the 6 counterterms power series
then the power series expansion of any Schwinger function of the model expressed
in powers of the renormalized couplings has a finite limit at all orders. More precisely

\begin{theorem}
\label{mainren}
There exist 6 counterterms $  CT^{\Lambda}_{6;1}, CT^{\Lambda}_{6;2}, CT^{\Lambda}_{4;1}, CT^{\Lambda}_{4;2}, CT^{\Lambda}_{2;1},  CT^{\Lambda}_{2;2}$, each of which a multi-power series of the four renormalized couplings
$(  \lambda^{\text{ren}}_{6;1}, \lambda^{\text{ren}}_{6;2}, \lambda^{\text{ren}}_{4;1}, \lambda^{\text{ren}}_{4;2})$, with $\Lambda$ dependent coefficients, such that
$CT^{\Lambda}_{6;1}, CT^{\Lambda}_{6;2}, CT^{\Lambda}_{4;1}, CT^{\Lambda}_{4;2}$ have  valuation at least 2  and $CT^{\Lambda}_{2;1}, CT^{\Lambda}_{2;2}$ have
valuation at least 1, and such that if the bare couplings in equation are defined as
\bea   \lambda^{\Lambda}_{6;1} &=&  \lambda^{\text{ren}}_{6;1}  + CT^{\Lambda}_{6;1} (  \lambda^{\text{ren}}_{6;1}, \lambda^{\text{ren}}_{6;2}, \lambda^{\text{ren}}_{4;1}, \lambda^{\text{ren}}_{4;2}) \\
   \lambda^{\Lambda}_{6;2} &=&  \lambda^{\text{ren}}_{6;2}  + CT^{\Lambda}_{6;2} (  \lambda^{\text{ren}}_{6;1}, \lambda^{\text{ren}}_{6;2}, \lambda^{\text{ren}}_{4;1}, \lambda^{\text{ren}}_{4;2}) \\
   \lambda^{\Lambda}_{4;1} &=&  \lambda^{\text{ren}}_{4;1}  + CT^{\Lambda}_{4;1} (  \lambda^{\text{ren}}_{6;1}, \lambda^{\text{ren}}_{6;2}, \lambda^{\text{ren}}_{4;1}, \lambda^{\text{ren}}_{4;2}) \\
   \lambda^{\Lambda}_{4;2} &=&  \lambda^{\text{ren}}_{4;2}  + CT^{\Lambda}_{4;2} (  \lambda^{\text{ren}}_{6;1}, \lambda^{\text{ren}}_{6;2}, \lambda^{\text{ren}}_{4;1}, \lambda^{\text{ren}}_{4;2}) 
\eea
then the Schwinger functions of the model with partition function \ref{baremodel},   when re-expressed as multi-power series in the four renormalized couplings
$( \lambda^{\text{ren}}_{6;1}, \lambda^{\text{ren}}_{6;2}, \lambda^{\text{ren}}_{4;1}, \lambda^{\text{ren}}_{4;2}) $,
have all their coefficients finite when the ultraviolet cutoff $\Lambda$ goes to infinity.
\end{theorem}
This is the usual statement of perturbative renormalizability of the model.
The rest of the paper is devoted to the proof of this theorem. 

\section{Multiscale Analysis}
\label{sect:multiscal}

In this section, we define the multiscale analysis
leading to the power counting and proof of Theorem \ref{mainren}.
First, we need a bound on the propagator
adapted to the scale analysis. Then, we apply the usual multiscale 
formalism  \cite{Rivasseau:1991ub}. We obtain a prime power 
counting of the amplitude of any graph in term of its
high or quasi-local subgraphs. 
The fine analysis of this power counting and renormalization 
program will be differed  to the next sections.

\subsection{Decomposition and bounds on the propagator} 
\label{subsect:propabound}

Let us consider the $U(1)$ tensor dynamical model defined 
by the kinetic term 
\bea
\widehat{S}^{\,\kin} =\int_{h_{j}}
\bar\varphi_{1,2,3,4} \Big(-\sum_{s=1}^4 \Delta_{s} + m^2 \Big) \varphi_{1,2,3,4}\;.
\eea
Given a set of integers 
$(\{q_s\};\{q'_s\}):= (\{q_1,q_2,q_3,q_4\};\{q'_1,q'_2,q'_3,q'_4\})$,
$q_s,q_s'\in \mathbb{Z}$, the kernel of the propagator 
$[-\sum_{s=1}^4 \Delta_{s} + m^2]^{-1}$ in momentum space can be written
\bea
C(\{q_s\};\{q'_s\})
= \Big[\sum_{s=1}^4 (q_{s})^2 + m^2\Big]^{-1}  [\prod_{s=1}^4 \delta_{q_s,q'_s} ]\;.
\label{kern}
\eea
Choosing a local coordinate system on  $S^1\sim U(1)$, 
parameterized by $\theta \in [0,2\pi)$, consider a set 
of such coordinates $(\{\theta_{s}\}; \{\theta'_s\}) 
:= (\{\theta_{1},\theta_{2},\theta_{3},\theta_{4}\}; \{\theta'_{1},\theta'_{2},\theta'_{3},\theta'_{4}\})$ and the corresponding elements
$(\{h_{s}\}; \{h'_s\})$ such that 
$h_{s}= e^{i q_s \theta_s}$, $s=1,\dots,4$. 
The kernel (\ref{kern}) expressed in the direct space can be
evaluated as
\beq
C(\{\theta_{s}\}; \{\theta'_s\})= \sum_{q_s,q'_s \in \mathbb{Z}} 
C(\{q_s\};\{q'_s\})  e^{i \sum_{s} [q_s \theta_s - q'_s\theta'_s]}
=\sum_{q_s\in \mathbb{Z}}
\int_0^{\infty} e^{-\alpha   \left[ \sum_{s} q_{s}^2 + m^2\right] 
+ i \sum_s q_s (\theta_s -  \theta'_{s})  } d\alpha\;, 
\label{param}
\eeq
where we have introduced a Schwinger parameter $\alpha$. 

A direct calculation yields, up to some unessential constant
$k=\pi^2$
\bea
C(\{\theta_{s}\}; \{\theta'_s\}) &=& k
\int_0^{\infty} \frac{e^{-m^2\alpha  }}{\alpha ^2}
e^{ - \frac{1}{4\alpha }\sum_s  [\theta_s -  \theta'_{s}]^2 } \;
 T(\alpha;\{\theta_{s}\}; \{\theta'_s\})  \; d\alpha\;,\crcr
T(\alpha;\{\theta_{s}\}; \{\theta'_s\})
&=&\prod_{s=1}^4 \left\{ 
1+ 2\sum_{n=1}^\infty e^{-\frac{\pi^2n^2}{\alpha}} \cosh\Big[  \frac{n\pi}{\alpha}[\theta_s -  \theta'_{s}]\Big]
\right\} \;,
\label{propinit}
\eea
where $T$ can be related to 
 the third  Jacobi elliptic function (although this special function
is not used in this paper) \cite{abram}. This is the general 
expression of the covariance in this $U(1)$ theory, which is the simplest finite volume four-dimensional theory with periodic
boundary conditions on the Laplacian. 
The latter is an important feature of this theory: 
amplitudes and functions involving the quantities $|\theta_{s}-\theta'_{s}|$ will be all translation invariant as functions on the torus.

In the following developments, we do not actually 
need the explicit expression of this propagator but
only its behavior at small distance will be useful.
The problem of infrared divergences is simply avoided in this paper
by the fact that $U(1)$ is compact and for simplicity we 
can even assume $m^2= 0$, to have no problem with the zero mode
of the propagator.

We introduce the usual slice decomposition of the propagator:
\bea\label{scaledec}
&&
C  = \sum_{i=0}^{\infty} C_i \;,
\crcr 
&&
C_0(\{\theta_s\}; \{\theta_s'\}) =
  k\int_{1}^{\infty} \frac{e^{-m^2\alpha  }}{\alpha ^2}
e^{ - \frac{1}{4\alpha }\sum_s  [\theta_s -  \theta'_{s}]^2 } \,
T(\alpha;\{\theta_s\}; \{\theta_s'\})\, d\alpha \;,
\crcr
&&
C_i(\{\theta_s\}; \{\theta_s'\}) = 
k \int^{M^{-2i}}_{M^{-2(i+1)}} 
\frac{e^{-m^2\alpha  }}{\alpha ^2}
e^{ - \frac{1}{4\alpha }\sum_s  [\theta_s -  \theta'_{s}]^2 } \;
T(\alpha;\{\theta_s\}; \{\theta_s'\}) \,  d\alpha\;.
\eea
The following statement holds
\begin{lemma}
\label{lem1}
For all $i=0,1,\dots$, for all $\um \in \N$ there exist some constants $K\geq 0$, $K_{\um}\geq 0$ and $\delta \geq 0$ such that
\bea
&&
C_i(\{\theta_s\}; \{\theta_s'\})  \leq 
K M^{2i} e^{-\delta M^{i} \sum_{s=1}^4|\theta_s -  \theta'_{s}|} \;, 
\label{boundprop} \\
&&
\Big(\prod_{k=1}^\um \partial_{\theta',s_k}\Big) C_i(\{\theta_s\}; \{\theta_s'\})  \leq 
K_\um M^{(2+\um)i} e^{-\delta M^{i} \sum_{s=1}^4|\theta_{s} -  \theta'_{s}|} \;.
\label{boundder}
\eea
\end{lemma}
\noindent{\bf Proof.}
For small $\alpha$,  the propagator can be faithfully approximated by a heat kernel.  In a high slice $i\gg 1$,  the following bound is valid (see Appendix 
\ref{app:prop} for the calculation details)
\beq
C_i(\{\theta_s\}; \{\theta_s'\}) \leq 
K' \int^{M^{-2i}}_{M^{-2(i+1)}} d\alpha \;\frac{e^{-m^2\alpha }}{\alpha^2}
e^{ - \frac{1}{4\alpha}\sum_s  [\theta_s -  \theta'_{s}]^2 }
\leq K'' M^{2i} e^{-\delta' M^{2i}\sum_s |\theta_s -  \theta'_{s}|^2} \;,
\label{ci}
\eeq
where $K'$, $K''$ and $\delta'$ are  constants. From this last expression, 
we get the useful bound
\beq
C_i(\{\theta_s\}; \{\theta_s'\})  \leq 
K M^{2i} e^{-\delta M^{i} \sum_{s}|\theta_s -  \theta'_{s}|} \;,
\eeq
with $K$ and $\delta$ some constants and 
the sum is performed over $s=1,\dots,4$. For the last slice,
we have (see Appendix \ref{app:prop})
\bea
C_0(\{\theta_s\}; \{\theta_s'\}) &\leq&  K' \int_{1}^{\infty} \;\frac{e^{-m^2\alpha/2 }}{\alpha^2}  d\alpha 
\leq K e^{-\delta\sum_s |\theta_s -  \theta'_{s}|} \;,
\label{coin}
\eea
where we used the fact that $ |\theta_s -  \theta'_{s}| \le 2 \pi  $. 
This proves the first bound (\ref{boundprop}). 

For the second inequality, we can differentiate $\um$ times the propagator with respect to a set of strands and get  (note
that we reintroduce the momentum space representation as
intermediate step using a slice decomposition from \eqref{param} for simplifications)
\bea
&&
 |\Big(\prod_{k=1}^\um\partial_{\theta',s_k}\Big) C_i(\{\theta_s\}; \{\theta_s'\})| 
\leq \crcr
&& 
|\sum_{q_s\in \mathbb{Z}} \int^{M^{-2i}}_{M^{-2(i+1)}} 
\frac{1}{\alpha^{\um/2}}\big[\prod_{k=1}^\um(-i\sqrt{\alpha}q_{s'_k}) \big] e^{-\alpha[\sum_{s} q_s^2 + m^2]  } 
e^{  i\sum_s  q_s[\theta_s -  \theta'_{s}]} \;
 \,  d\alpha |\\
&&   \leq  K'_\um |\sum_{q_s\in \mathbb{Z}} \int^{M^{-2i}}_{M^{-2(i+1)}} 
\frac{1}{\alpha^{\um/2}} e^{-\alpha[\sum_{s} q_s^2 + m^2]  } 
e^{  i\sum_s  q_s[\theta_s -  \theta'_{s}]} \;
 \,  d\alpha | \leq K_\um M^{2i +2(\um/2) i} e^{-\delta M^{i}\sum_{s}|\theta_s -  \theta'_{s}|}\;,
\nonumber
\eea
 where, in the last stage, we perform the summation in $q$'s
and use again the bound on $T$ (according to the same procedure
yielding \eqref{ci}).

Similarly, for the last slice, we have
\bea
|\Big(\prod_{k=1}^\um\partial_{\theta',s_k}\Big) C_0(\{\theta_s\}; \{\theta_s'\})
| &\leq& K'
 \int_{1}^{\infty} \;\frac{e^{-m^2\alpha/2 }}{\alpha^{2 + \um/2}}  d\alpha 
\leq K_m  e^{-\delta\sum_s |\theta_s -  \theta'_{s}|} \;.
\eea

\qed

Imposing an ultraviolet cutoff consists in summing the slice index only up to a large integer 
$\Lambda$ in \eqref{scaledec}
\beq
C^{\Lambda}  = \sum_{i=0}^{\Lambda} C_i \; ,
\eeq
and the ultraviolet limit is $\Lambda\to\infty.$ From now on, we forget the superscript $\Lambda$
most of the time for simplicity.

\subsection{Momentum attributions and optimal amplitude bound}
\label{subsect:model}

Let us consider a connected amputated graph $\cG$ with set of vertices 
 $\cV$, with cardinal $V= |\cV|$, and $\cL$ set of lines, with cardinal  $L= |\cL|$. Let $N_{\ext}$ be the number of external 
fields or  legs. Since  wave-function counterterms have special power counting
because they carry an extra $p^2$,
we suppose first for simplicity  that the graph does not have wave-function counterterm, then we add the easy
correction for wave function counterterms.

\medskip
\noindent{\bf Direct space -}
 The bare amplitude associated with $\cG$ is of the form
\beq
A_{\cG} = \sum_{\mu} \int [\prod_{v,s} d\theta_{v, s}] 
[\prod_{\ell \in \cL} C_{i_\ell(\mu)}
(\{\theta_{v, \ell(v), s}\};\{\theta_{v',\ell(v'), s}\})][
\prod_{v\in \cV; \;s}  \delta(\theta_{v, s}-\theta_{v, s'})]\;,
\eeq
where $\theta_{v, \ell(v), s}$ are coordinates involved
in the propagator which 
should possess a vertex label $v$, a strand label $s$ 
but also a line index $i_\ell$;  $\theta_{v s}$ are 
the same position coordinates involved in the vertex
which should have both vertex $v$ and strand $s$ labels;
$\delta(\theta_{v, s}-\theta_{v, s'})$ is the delta Dirac 
distribution on the torus;
$\mu= (i_1,i_2,\dots,i_q)$ is a multi-index called momentum 
assignment which gives to each propagator of each internal line 
$\ell$ of the graph  a scale $i_\ell  \in [0,  \Lambda ]$; the sum over $\mu$ is performed on all possible assignments. The graph being amputated, there is no 
external propagator but rather external vertices where test 
functions or external fields can be hooked. 
It is conventional to give a fixed scale $i_{\ext}=-1$ for those external lines. We focus on $A_{\cG;\mu}$. The sum $A_{\cG}=\sum_{\mu} A_{\cG;\mu}$
can be done only after renormalization.

The next stage is to perform some spatial integrations of the $\theta_{v,s}$ vertex variables in $A_{\cG;\mu}$. The main point is to bound this integral 
in an ``optimal'' way.

Given a momentum assignment $\mu$ and a fixed scale $i$, we consider the complete list
of the connected components $G^{(k)}_i$, $k=1,2,\dots,k(i)$, 
of the subgraph $\cG_i$ made of all lines in $\cG$
with the scale attribution $j\geq i$ in $\mu$. These subgraphs called high or quasi-local 
are the key objects in the multiscale expansion  \cite{Rivasseau:1991ub}.
A partial (inclusion) order can be defined on the set of $G^{(k)}_i$
and $\cG_0=\cG$. The abstract tree 
made of nodes as the $G^{(k)}_i$ associated to that partial order
is called the Gallavotti-Nicol\`o tree \cite{Galla}, for which 
$\cG$ is merely the root. Given an arbitrary subgraph $g$, one defines:
\beq
i_g(\mu) = \inf_{l\in g}i_l(\mu) \;,  \qquad e_g(\mu) = 
\sup_{l \,\text{external line of}\, g} i_l(\mu) \;.
\eeq
The first quantity is the lowest scale inside the subgraph $g$
whereas the second corresponds to the higher scale of all
lines (external to $g$) to which the subgraph $g$ is hooked.
The subgraph $g$ is a $G^{(k)}_i$ for a given $\mu$ 
if and only if $i_g(\mu) \ge i >e_g(\mu) $, in other words,
any internal scale is higher than the greater external scale. 
In the ordinary field theory situation, 
the key point is to optimize the bound over spatial integrations 
by choosing   a spanning tree $T$ of $\cG$ \footnote{ A spanning tree of $\cG$ is a set of lines passing through all vertices of $\cG$ 
without forming loop. Integrations on vertex variables 
will be associated with the choice of $T$.} compatible with the abstract
Gallavotti-Nicol\`o tree. This can be done by considering
the restriction $T_i^k$ of $T$ to any $G^{(k)}_i$ in such way 
that $T_i^k$ is  still a spanning tree for $G^{(k)}_i$. 

The present situation is slightly different. 
Due to the particular form of the vertex operator,
i.e. a product of delta functions, the graph amplitude $A_{\cG}$ factorizes
in term of closed or open circuit called ``faces''. 
This notion coincides with ribbon graph faces in matrix model, see
for instance \cite{Rivasseau:2007ab}.  Let $\cF$ be the set of such faces which 
can be decomposed in closed or internal faces, say $\cF_{\inter}$,
and open faces which touch on external vertices, we call them
$\cF_{\ext}$. The cardinal $F$ of $\cF$ is of course 
the sum of $F_{\ext}$ and $F_{\inter}$ cardinal of $\cF_{\inter}$
and $\cF_{\ext}$, respectively. 

We write, using at first  the bound (\ref{boundprop}) (and dropping
some indices, for simplicity $i_\ell(\mu)=i_\ell$),
\bea
|A_{\cG;\mu}| &\leq& 
 \int [\prod_{v} d\theta_{v, s}] 
[\prod_{\ell \in \cL} 
K M^{2i_\ell} e^{-\delta M^{i_\ell } \sum_{s=1}^4|\theta_{v,i_\ell ,s} -  \theta'_{v,i_\ell, s}|} ]
\prod_{v\in \cV}  \delta(\theta_{v, s}-\theta_{v, s'}) \crcr
&\leq&
[\prod_{\ell \in \cL} 
K M^{2i_\ell} ]
\int [\prod_{f\in \cF}\prod_{ \ell \in f }  d\theta_{\ell,f}]  
\prod_{f\in \cF} \prod_{ \ell \in f } 
e^{-\delta M^{i_\ell } |\theta_{\ell, f }-\theta'_{\ell, f}|} \;.
\label{aprox}
\eea
In the last step, we simply rewrite the amplitude in terms
of faces and the product $\prod_{\ell \in f}$ is performed 
over all lines (here strands)  from which a given face $f$ is built.
 After this factorization, each variable 
$\theta$ can be now  indexed by a couple $(\ell,f)$, i.e.
by a unique face and a line where it can appear. Each 
variable can only appear (at most) in two such lines.

\begin{figure}
 \centering
     \begin{minipage}[t]{.8\textwidth}
      \centering
\includegraphics[angle=0, width=14cm, height=6cm]{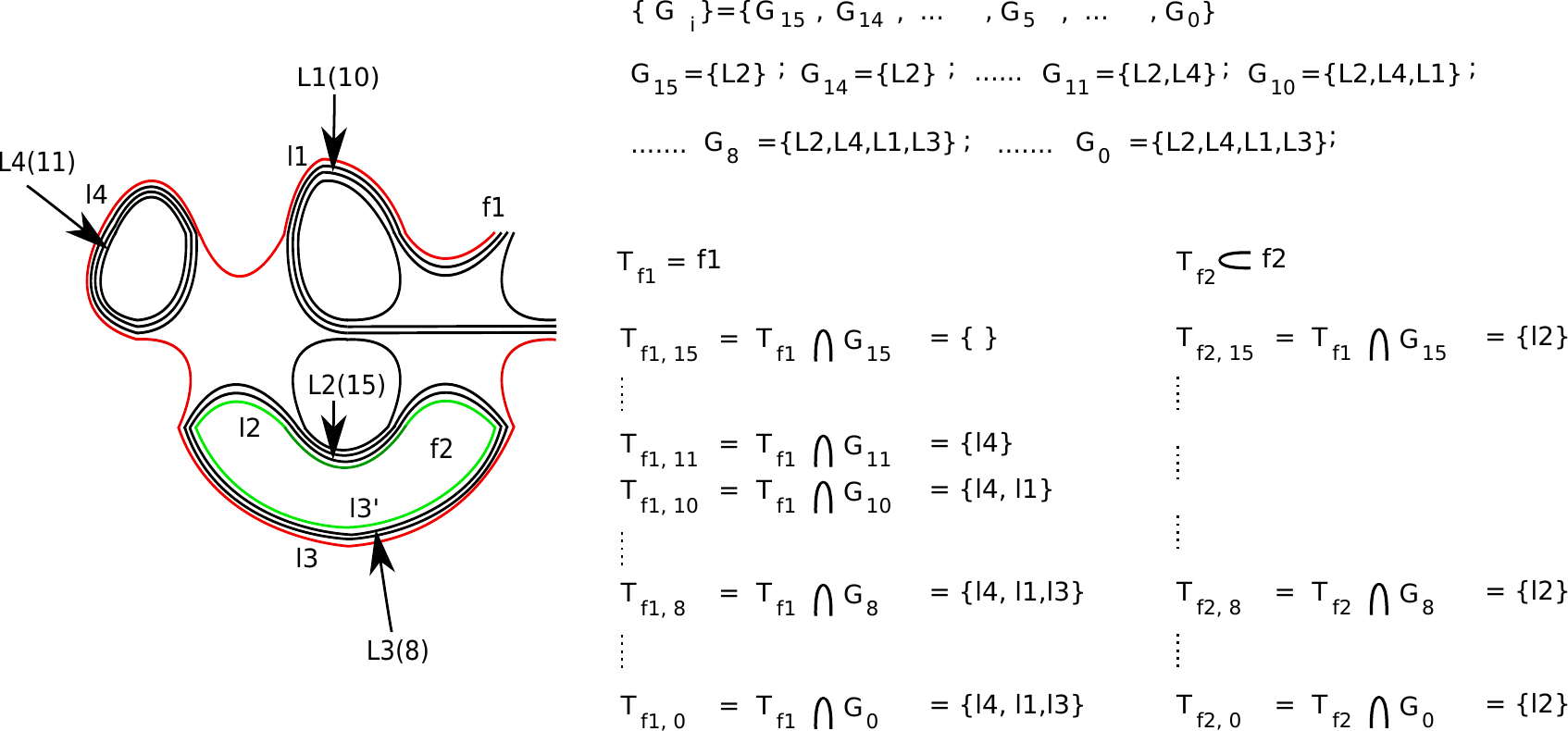}
\vspace{0.1cm}
\caption{ {\small A graph with a given scale attribution: 
lines of the graph are $\{L1,L2,L3,L4\}$ with 
scale $\{10,15,8,11\}$; the face $f1$ (in red) is open
and formed by the strands $l1,l4,l3$; the face $f2$ (in green) is closed
and formed by $l2,l3'$. All $G^{(k=1)}_i$ have a unique 
connected component. The trees $T_{f1}$ and $T_{f2}$
decomposed on scales appear as $T_{f1, i}$ and $T_{f2, i}$
the set of which forms a spanning forest of $f1$  and $f2$,
respectively.
\label{fig:trees}}}
\end{minipage}
\end{figure}

Each integration in position coordinates $\theta$ 
will bring a ``good'' factor of $M^{-i}$. We therefore
need to integrate as much as possible position coordinates with decay factors from
high indices $i_\ell$  in order to bring more convergence. There is a way to optimize the bound on position integrations but, first, 
let us define the ``scale of a strand'' as the same scale of the 
line generating this strand.  We can integrate all positions 
but one in each open or closed faces. 
Along any open face of a $G^{(k)}_i$, we can integrate
all positions but one; each integration will give
a factor $M^{-i_\ell}$ corresponding to the 
strand (and its scale index) where the integrated variable belongs. It remains a 
last integration with respect to a variable touching an external strand. It
will be made later with a lower line and that will bring a larger factor of $M^{-j}$,
with $j\le i_\ell -1$. Hence for an open face, all internal decays
can be used once. For closed faces, 
the problem is similar but one last integration cannot 
be performed with scaled decays. The integration of the closed face 
can be optimized by simply choosing the highest
scales of strands belonging to the face. This integration,
similar to a momentum routine, will be performed
on position labels on a tree $T_f$  (this tree 
is not a tree of lines as in ordinary quantum field theory, but a tree
made of strands which possess also a scale index) associated with the face $f$.
If $f$ is open $T_f=f$, if $f$ is closed $T_f\subsetneq f$
and $T_f$ consists in the set of all strands of $f$ save one. 
The tree $T_f \subseteq f$
(which is the analog of the spanning tree $T$) will be
chosen to be compatible with the abstract Gallavoti-Nicol\`o 
tree associated with the $G^{(k)}_i$ in the sense that, 
the restriction $T_{f,i}^k= T_f\cap G^{(k)}_i$ is an open face in $G^{(k)}_i$ (see Fig. \ref{fig:trees}). 
The set of $T_{f,i}^k$ is called a ``spanning forest'' and is made of
a set of connected strands, a tree, belonging to the same 
face.

We rewrite each factor in terms of $G^{(k)}_i$ in the manner of \cite{Rivasseau:1991ub}:
\bea
&&
\prod_{\ell \in \cG} M^{2i_\ell} = \prod_{\ell \in \cG} \prod_{i=1}^{i_\ell} M^{2} = 
\prod_{\ell \in \cG}\prod_{(i,k)\in \mathbb{N}^2 / \ell \in G^{(k)}_i} M^2 =\prod_{(i,k)\in \mathbb{N}^2}  \prod_{\ell \in G^{(k)}_i} M^2 =
\prod_{(i,k)\in \mathbb{N}^2}M^{ 2L (G^{(k)}_i)}\;; \cr\cr
&&
[\prod_{ f \in \cF_{\ext} }   \prod_{ \ell \in f }  M^{ - i_\ell }]
[\prod_{ f \in \cF_{\inter}}  \prod_{ \ell \in T_f \subset f }  M^{ - i_\ell }]
=[\prod_{ f \in \cF_{\ext} }\prod_{\ell \in f} \prod_{i=1}^{i_\ell} M^{-1}]
 [\prod_{ f \in \cF_{\inter}}  \prod_{ \ell \in T_f \subset f }   \prod_{i=1}^{i_\ell} M^{-1}] \cr\cr
&& 
 = [\prod_{  f\in \cF_{\ext}} \prod_{ \ell \in f }
\prod_{(i,k)\in \mathbb{N}^2 / \ell \in G^{(k)}_i} 
M^{-1} ]
 [\prod_{ f \in \cF_{\inter}}  \prod_{ \ell \in T_f \subset f }   \prod_{(i,k)\in \mathbb{N}^2 / \ell \in G^{(k)}_i}  M^{-1}] \crcr
&& =  \prod_{(i,k)} \big[ 
 [\prod_{  f\in \cF_{\ext }  \cap  G^{(k)}_i } \;
\prod_{ \ell \in  T^k_{f,i}= f \cap  G^{(k)}_i }  M^{-1} ]
 [\prod_{ f \in \cF_{\inter} \cap  G^{(k)}_i } \;
\prod_{ \ell \in  T^k_{f,i}= T_f \cap  G^{(k)}_i }  M^{-1}] \big]\crcr
&&
= \prod_{(i,k) }\; \prod_{  f\in \cF \cap  G^{(k)}_i}  
\prod_{ \ell \in  T^k_{f,i} }  M^{-1}  =  \prod_{(i,k)}
M^{-4 L(G^{(k)}_i) + F_{\inter}(G^{(k)}_i)} \;,
\label{res}
\eea
where $ L(G^{(k)}_i) $ and $ F_{\inter}(G^{(k)}_i)$ denote the 
number of internal lines and internal faces of the subgraph $G^{(k)}_i$, respectively.
In the last equality, we use the fact that, given line at scale $i$,
all $4$ strand positions can be integrated
and then they should contribute to the same 
the subgraph $G^{(k)}_{i}$. However, since one position 
label per closed face is not integrated (with any line decay) but is simply integrated over the torus (without decay), it brings a full factor 1. Therefore we have the relation
\bea
4 L(G^{(k)}_{i}) = \sum_{ f\in \cF_{\ext } \cap  G^{(k)}_i } |T^k_{f,i}| + 
\sum_{f\in \cF_{\inter }  \cap  G^{(k)}_i}  (|T^k_{f,i}| + 1 ) \;,
\qquad 
F_{\inter} (G^{(k)}_i) = \sum_{f\in \cF_{\inter }  \cap  G^{(k)}_i}  1 \;.
\eea
Combining the results (\ref{res}) with the factors coming from
spatial integrations, we obtain a bound of the
graph amplitude at a given attribution $\mu$
\bea  \label{goodbound}
|A_{\cG;\mu}| \leq K^{n} \prod_{(i,k) }
M^{-2 L(G^{(k)}_i) + F_{\inter}(G^{(k)}_i)} \;,
\eea
where $K$ is some constant and $n$ is the number of vertices of the graph (assumed without wave-function counterterms). 

\medskip
\noindent{\bf Momentum basis -}
We briefly sketch in this paragraph how
the same prime power counting can be recovered in the 
momentum basis.  The action will mainly remain
 the same: each field has to be replaced
by a tensor $\varphi_{m_1,m_2,m_3,m_4}$
and the interaction pattern will be the same 
but with respect to the discrete indices $m_{i}\in \mathbb{Z}$.
In this representation, the propagator
kernel is of the form (\ref{kern}) and we can use 
its parametric form. 
Decomposing the propagator in the same  slices $C = \sum_{i} C_i $, 
we have
\bea
C_{i}(\{q_s\};\{q'_s\}) &=& \int_{M^{-2(i+1)}}^{M^{-2i}}\; d\alpha\;
 e^{-\alpha [(q_{1})^2+(q_{2})^2+(q_{3})^2+(q_{4})^2 + m^2]}
\prod_{s=1}^4 \delta_{q_s, q'_s}
\crcr
& \leq & K M^{-2i} e^{-\delta  M^{-i}[\sum_{s} |q_{s}| + m^2]}
\prod_{s=1}^4 \delta_{q_s, q'_s} \;;
\crcr
C_1(\{q_s\};\{q'_s\}) &=& \int_{1}^{\infty}\; d\alpha\;
 e^{-\alpha [(q_{1})^2+(q_{2})^2+(q_{3})^2+(q_{4})^2 + m^2]}
\prod_{s=1}^4 \delta_{q_s, q'_s} \leq K \prod_{s=1}^4 \delta_{q_s, q'_s}\;.
\label{mombound}
\eea
Given a momentum assignment $\mu$,
the multiscale representation of a given amputated graph amplitude can be expressed as
\bea
A_{\cG;\mu}= 
\sum_{q_{v,s}}\; \prod_{\ell \in \cL} C_{i_\ell(\mu)}(\{q_{v,\ell(v),s}\};\{q_{v',\ell(v'),s}\})
\prod_{v\in \cV; s} \delta_{q_{v, s}, q_{v, s '}}\;,
\label{amplimom}
\eea
the last $\delta$'s are Kronecker symbols associated to vertices. 
The sum is performed on all integers $q_{v,s} \in \mathbb{Z}$ 
 in the momentum basis. 
Using the fact that faces factor in the amplitude, 
we obtain
\bea
|A_{\cG;\mu}|
&\leq&  K^n 
\prod_{\ell \in \cL}  M^{-2i_\ell } 
\sum_{q_{s}}
\prod_{\ell \in \cL} \prod_{s=1}^4 \delta_{q_{i_\ell s}, q'_{i_\ell s}}
 e^{-\delta  M^{-i_\ell}[\sum_{s} |q_{s}| + m^2]} \crcr
&\leq&
 K^n 
\prod_{\ell \in \cL}  M^{-2i_\ell } 
\sum_{q_{f}}
\prod_{f \in \cF} \prod_{\ell \in f}
e^{-\delta  M^{-i_\ell}  |q_{f}|} \;,
\eea
where the bound (\ref{mombound}) has been used. 
We introduced also $q_f$ as momenta per face amplitude
and the notation ``$l \in f$'' to mention the particular line 
(in fact, strand) contributing to the face $f$.
Two cases may occur: (1) the face $f \in \cF_{\inter}$, then the face amplitude is of the form 
$\sum_{q_f} e^{-\sum_{\ell \in f} \delta M^{-i_\ell} |q_{f}|}$.
Presently, we optimize by taking the lowest possible $i_\ell$ in the 
face because, up to some constants $\delta,\delta'$, 
$\sum_{p \in \mathbb{N} } 
e^{-\delta M^{-i} p }  =  \delta' M^{i} + O( M^{-i})$;
(2) the face $f$ is open, then all sums in $q_s$
can be performed and one gets $O(1)$. 
Hence, the first step is to bound again the above amplitude
by only terms involving $\cF_{\inter}$. 
Its remains to choose a tree $T_f$ of each internal face $f\in \cF_{\inter}$
compatible with the Gallavoti-Nicol\`o tree in the same
way as done above. The contributions can be again 
recast in terms of the $G^{(k)}_i$ (once again by introducing the restriction
of $T^k_{f,i} = T_f \cap G^{(k)}_i$). One infers  the same 
bound as given by \eqref{goodbound}.

The previous bounds applies to connected graphs without wave function counterterm. If the graph contains
such counterterm  vertices of the type $S_{2;2}$, let  $V'_2$ be their number. For each such counterterm,
we have an extra $p^2$ hence $M^{2i} $ factor. Hence, we finally have:

\begin{lemma}
\label{primepow}
For a connected graph $\cG$ (with external arguments integrated
versus fixed smooth test functions), we have
\beq
|A_{\cG;\mu}| \leq K^n  \prod_{(i,k)}
M^{ \omega_d(G^{(k)}_i)}\;,
\eeq
where $K$ and $n$ are large constants, 
$\omega_d(G^{(k)}_i)= -2 L(G^{(k)}_i) + F_{\inter}(G^{(k)}_i)  + 2 V'_2 (G^{(k)}_i)  $. 
\end{lemma}
We call degree of divergence of the graph $\cG$, the quantity
\bea
\omega_d(\cG) =  -2 L(\cG) +F(\cG)   + 2 V'_2(\cG)   \;.
\label{convergence}
\eea

\section{ Divergence Degree and Topology}
\label{sect:divdeg}

\begin{figure}
 \centering
     \begin{minipage}[t]{.8\textwidth}
      \centering
\includegraphics[angle=0, width=12cm, height=5cm]{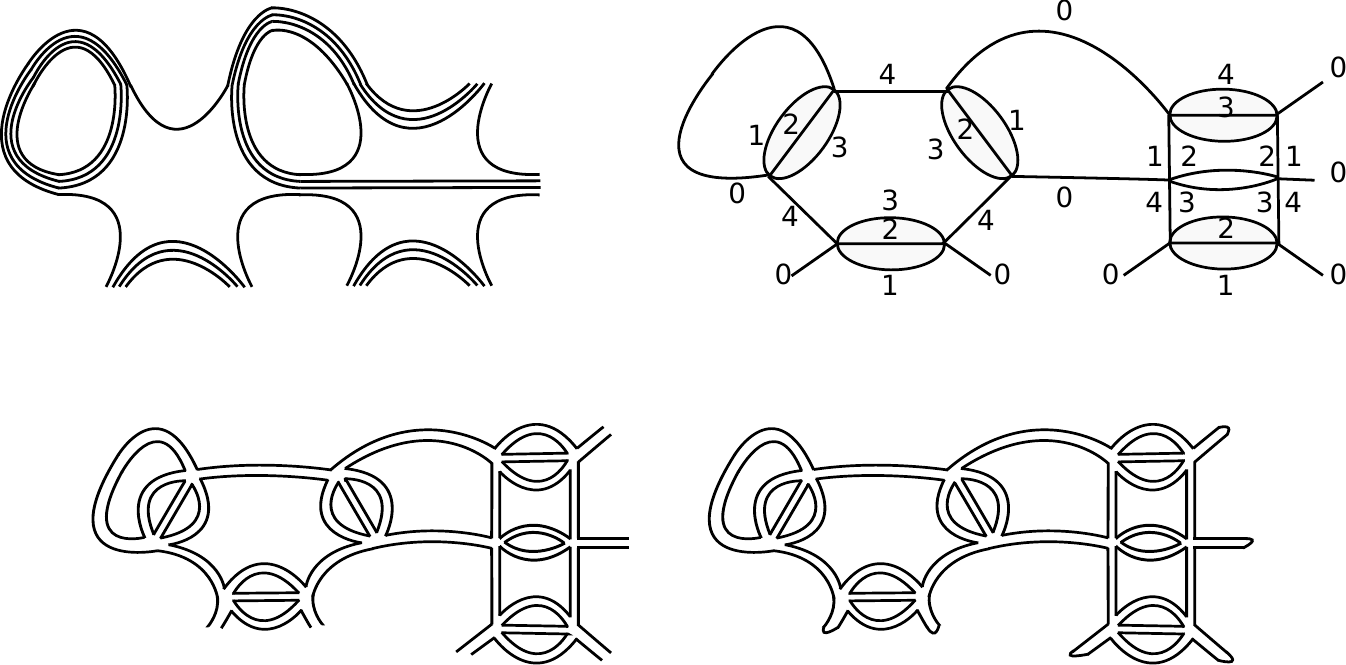}
\vspace{0.1cm}
\caption{ {\small A graph $\cG$, its colored extension $\cexG$ (five valence vertices with colored (half-)lines), 
the jacket subgraph $J$ $(01234)$ of $\cexG$ and its associated 
pinched jacket $\tJ$. \label{cc}}}
\end{minipage}
\put(-288,70){$\cG$}
\put(-110,-12){$\tJ$}
\put(-265,-12){$J$}
\put(-105,70){$\cexG$}
\end{figure}

\begin{figure}
 \centering
     \begin{minipage}[t]{.8\textwidth}
      \centering
\includegraphics[angle=0, width=6cm, height=2cm]{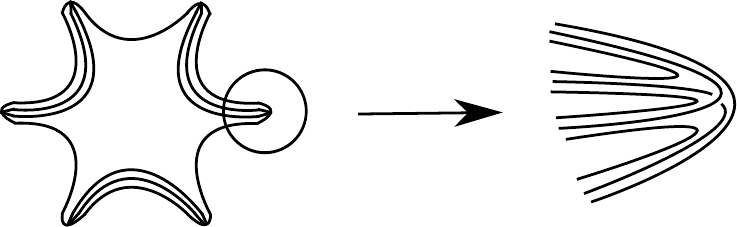}
\vspace{0.1cm}
\caption{ {\small The boundary $\bG$ of $\cG$ (see Fig.\ref{cc}) and its rank 3 
tensor structure.\label{cc2}}}
\end{minipage}
\end{figure}

This section establishes another expression 
for the divergence degree $\omega_d(G^{(k)}_i)$
in an adequate form for the renormalization procedure. 
We will consider a general graph $\cG$ rather than some $G^{(k)}_i$
and we introduce more ingredients for carrying through the analysis.

The following definitions follow the main ideas
of jackets \cite{sefu3,Gur3,GurRiv} and boundary graph  \cite{Gurau:2009tz}.

\begin{definition}
Let $\cG$ be a graph in our $4$ dimensional theory. 
\begin{enumerate}
\item[(i)] We call colored extension of $\cG$ the unique  graph $\cexG$
obtained after restoring in $\cG$ the former colored theory graph
(see Fig.\ref{cc}).

\item[(ii)]  A jacket $J$ of $\cexG$ is a ribbon subgraph of 
$\cexG$ defined by a cycle $(0abcd)$ up to a cyclic permutation  (see Fig.\ref{cc}). There are 12
such jackets in dimension 4 \cite{GurRiv}. 

\item[(iii)] The jacket $\tJ$ is the jacket obtained from 
$J$ after ``pinching'' viz. the procedure consisting in 
closing all external legs present in $J$ (see Fig.\ref{cc}). 
Hence it is always a vacuum graph. 

\item[(iv)]  The boundary $\bG$ of the graph $\cG$
is the closed graph defined by vertices corresponding to 
external legs and by lines corresponding to external strands of $\cG$ 
\cite{Gurau:2009tz} (see Fig.\ref{cc2}). It is in our case a vacuum graph
of the $3$ dimensional colored theory.

\item[(v)]  A boundary jacket $\bJ$ is a jacket of 
$\bG$. There are 3 such boundary jackets in our case.

\end{enumerate}
\end{definition}

Consider a connected graph $\cG$.
Let $V_6$ be its number of $\int\varphi^6$ type vertices (of any type)
and $V_4$ its number of $\int\varphi^4$ vertices
of the type 1, $V_4'$ its number of vertices of 
type $(\int\varphi^2)^2$, $V_2$ the number of vertices of
the type $\int\varphi^2$ (mass counterterms) and $V'_2$
the number of vertices of the type $\int(\nabla\varphi)^2$ 
(wave function counterterms). Let $L$ be its number of lines and $N_{\ext}$ its number
of external legs. Consider also its colored extension $\cexG$
and its boundary $\bG$. 

Remark that the vertices contributing to $V_4'$ are
disconnected from the point of view of their strands. 
Hence it is convenient to reduce them in order to 
find the power counting with respect to only connected
component graphs. We will consider these types of 
vertices as a pair of two 2-point vertices $V_2''$, hence $V''_2 = 2V_4' $. 
The vertices $V_2''$ are identical to the mass vertices $V_2$ except that they occur
in pairs. The pairing is pictured in dotted line in Fig.3.
The power counting can be established separately for each
connected  component after removing all the dotted
lines.

The following statement, in the above notations, holds
\begin{theorem}
\label{convtop}
The divergence degree of a connected graph $\cG$ is an integer 
which writes
\bea  \label{contopformula}
\omega_d(\cG) = -\frac13 \left[ \sum_{J} g_{\tJ} 
-  \sum_{\bJ} g_{\bJ} \right] - (C_{\bG}-1) - V_4- 2(V_2 +V_2'') -  \frac12 \left[ N_{\ext}- 6\right],
\eea
where $g_{\tJ} $ and $g_{\bJ}$ are the genus of $\tJ$ 
and $\bJ$, respectively, $C_{\bG}$ is the number
of connected components of the boundary graph $\bG$; the first sum is performed on all closed jackets 
$\tJ$ of $\cexG$  and the second sum is performed on
all boundary jackets $\bJ$ of $\bG$.
\end{theorem}

\noindent{\bf Proof.} 
Given a connected graph (with respect to $V_2''$ and not to
$V'_4$) $\cG$ with the above characteristics,
we have the following relation between the numbers of 
lines, of external legs and of vertices:
\bea
6 V_6 + 4 V_4  +2 (V_2  + V'_2 + V_2'') = 2L + N_{\ext}\;.
\eea
Consider its colored extension $\cexG$. The latter graph is
connected. Its number of vertices $V_{\cexG}$ and
 its number of lines $L_{\cexG}$ satisfy
\beq
V_{\cexG} = 6 V_6 + 4 V_4  + 2 (V_2  + V_2' + V_2'')  \;, \quad
L_{\cexG} =  L + L_{\inter;\,\cexG} = \frac12 (5 V_{\cexG} - N_{\ext})\;,
\eeq
where $L_{\inter;\cexG}$ are the internal lines of  $\cexG$ which 
do not appear in $\cG$.
Let us denote $F_{\cexG}$ the number of faces of  $\cexG$. 
The latter can be decomposed as well in terms of the number of
faces of the initial graph, that is $F$, but also additional
faces $F_{\inter;\cexG}$ due to the internal colored structure:
\beq
F_{\cexG} = F + F_{\inter;\,\cexG} \;.
\eeq

Let us focus now on the 12 jackets of $\cexG$. Any such 
jacket is connected since $\cexG$ is connected. 
The following relations hold in the colored theory:
\bea
V_{J} = V_{\cexG} \;,\qquad 
L_{J} = L_{\cexG}\;, \qquad 
N_{\ext;\,J} = N_{\ext;\,\cexG} = N_{\ext}\;.
\eea
Like the initial graph, a jacket may have open and closed faces. 
Each face of the graph $\cexG$ (open or closed) is shared by exactly 
$(D-1)!=6$ jackets. We have
\bea
\sum_{J} F_{J} = 6 F_{\cexG}\;.
\eea

The Euler characteristic of an open ribbon graph (i.e. a ribbon 
graph with external legs) is not well defined. 
Nevertheless, closing all external half-lines in a ribbon 
graph leads to another unique (closed) ribbon graph for which the above topological
number is perfectly defined. This is the purpose of the pinching procedure
applied to $J$ leading to $\tJ$. The resulting jacket $\tJ$ has the same number of vertices, 
the same number of lines as $J$,  but a different number of faces than $J$. The number of faces of $\tJ$ can be partitioned into
$F_{\tJ} = F_{\inter; \tJ} + F_{\ext; \tJ}$, where $F_{\inter; \tJ}$
corresponds to  $F_{\inter; J} $ the number of faces of $J$
and $ F_{\ext; \tJ}$ is the number of additional closed faces
created by the pinching procedure. Using the formula for the Euler characteristics
of $\tJ$, we have
\beq
F_{\inter;\, \tJ} + F_{\ext;\, \tJ}= 2-2g_{\tJ} - V_{J} + L_{J}\; .
\label{eulertj}
\eeq
Note that all external pinched faces $\tJ$ come from some
open faces of the initial graph $\cG$. However $F_{\inter;\tJ}$ can be decomposed
in two categories of faces: one category of faces which belong to $\cG$ 
(the number of such faces is denoted by $F_{\inter;\,   \tJ; \, \cG}$)
and another category of faces belonging only to the internal structure of 
$\cexG$ (the number of these latter faces is denoted by $F_{\inter;\,  \tJ; \, \cexG}$)
Hence
\beq
F_{\inter;\,\tJ} = F_{\inter;\,   \tJ; \, \cG} + F_{\inter;\,   \tJ; \, \cexG}\;.
\eeq
The first step is to sum over all jackets
in the l.h.s of (\ref{eulertj}):
\beq
\sum_{J} ( F_{\inter;\,   \tJ; \, \cG} + F_{\inter;\,   \tJ; \, \cexG} + F_{\ext; \tJ}) = 
6 F_{\inter;\cG} + 6F_{\inter;\;  \cexG}  + \sum_{J} F_{\ext; \tJ}\;.
\eeq
Note that the number $F_{\inter;\;  \cexG}$  of internal faces 
of $\cexG$ can be directly evaluated from any graph: 
each $\varphi^6$ vertex contains
$12$ such faces whereas each $\varphi^4$ vertex contains $9$  and each
$\varphi^2$ type vertices contains $6$ internal faces
so that
\beq
F_{\inter;\; \cexG}  = 12 V_{6} +9 V_4+ 6(V_2 +  V_2' + V_2'')\;.
\eeq
Summing over all jackets the r.h.s. of (\ref{eulertj}), we focus on
the following part: 
\beq
\sum_{ J}\left[ - V_{J} + L_{J}\right]  
 =  12[9 V_6 + 6V_4  +3(V_2  +  V_2'  +  V_2'') ]- 6N_{\ext}\;. 
\eeq
Equating l.h.s and r.h.s, we extract the following 
relation for $F_{\inter;\cG}$:
\beq
 F_{\inter;\cG}  
=  - \frac16  \sum_{J} F_{\ext; \tJ}
- \frac13 \sum_{J} g_{\tJ} 
+4 +( 6V_6 + 3V_4)  - N_{\ext} \;.
\label{facint}
\eeq

\begin{figure}
 \centering
     \begin{minipage}[t]{.8\textwidth}
      \centering
\includegraphics[angle=0, width=14cm, height=3cm]{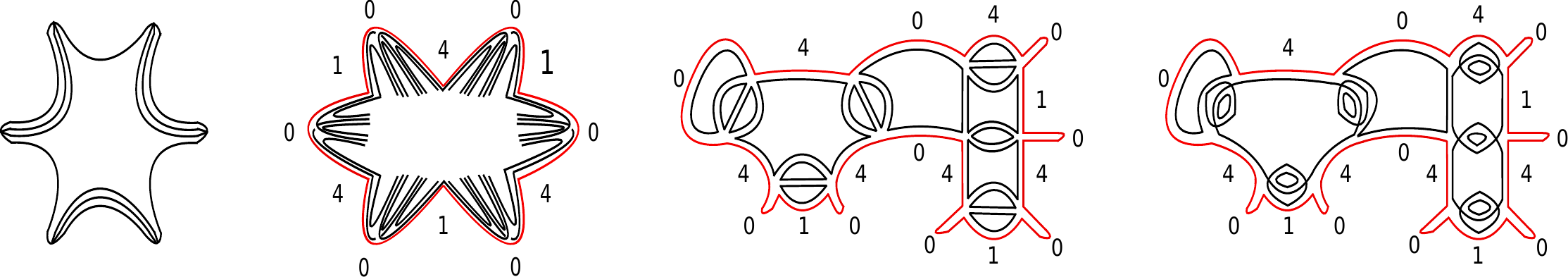}
\vspace{0.1cm}
\caption{ {\small The boundary $\bG$ of  graph $\cG$ (see Fig.\ref{cc}),
one of its colored face $f(014)$ (in red) and the unique two pinched jackets 
$\tJ(01234)$ and $\tJ'(01324)$ of $\cexG$ containing $f$ 
(as highlighted). 
\label{fig:boundface}}}
\end{minipage}
\put(-356,-12){$\bG$}
\put(-280,-12){$f$(014)}
\put(-155,-12){$\tJ$}
\put(-30,-12){$\tJ '$}
\end{figure}
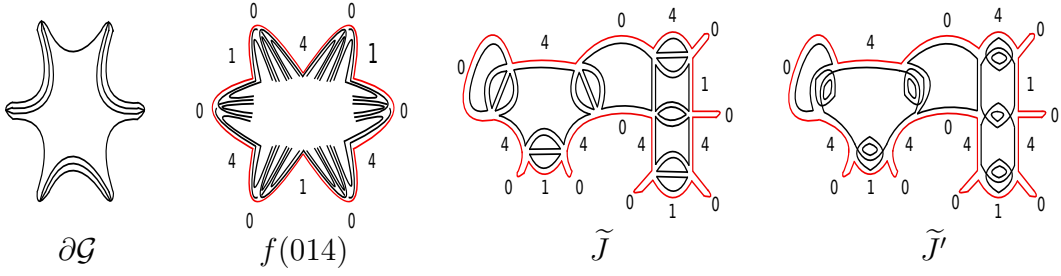

The next stage is to re-express $ \sum_{J} F_{\ext; \tJ}$
 in terms of topological numbers of the boundary graph  of the graph $\cG$. 
This boundary $\bG$  is defined such that 
\beq
V_{\bG} = N_{\ext} \;,\qquad
L_{\bG} = F_{\ext} \;.
\eeq
Since each external leg of the initial
graph $\cG$ has 4 strands and an external leg is made with 
two end-points belonging to two external legs, we have
\bea
4N_{\ext} =   2 F_{\ext}\;.
\eea
The boundary graph is a closed (vacuum) colored graph
living in the lower dimension $D-1 = 3$.
Hence, the boundary graph is again a tensor graph 
with jackets that will be denoted $\bJ$.
Remarkably, the boundary graph may be made of several connected
components. The degree of $\bG$ is defined as the sum of genera
of its jackets (which are all closed since $\bG$ is) which is
$\omega_{\bG} = \sum_{\bJ }g_{\bJ}$, 
where $g_{\bJ}$ is itself the sum of the genera
of its connected components labeled by $\rho$, i.e.
$
g_{\bJ}= \sum_{\rho}g_{{\bJ}_{\rho}}.
$
Naturally, some relations on the numbers of vertices 
and lines between the boundary graph and jackets can
be found:
\beq
V_{\bJ} = V_{\bG} = N_{\ext} \;,\qquad
L_{\bJ} = L_{\bG} = F_{\ext} \;.
\eeq
Let $F_{\bJ}$ the  number of faces of $\bJ$.
The ordinary three dimensional colored relations apply to $\bJ$ and $\bG$:
\beq
\sum_{\bJ} F_{\bJ} = (3-1)!  \;   F_{\bG} = 2  F_{\bG} \;, \qquad
\sum_{\bJ} 1 = \frac12 3! = 3\;.
\eeq
Thus,  we have, using the Euler characteristic formula for boundary jackets,
\beq
\sum_{\bJ} F_{\bJ}  = \sum_{\bJ} \left[  (2C_{\bJ}-2g_{\bJ})  - V_{\bJ} +L_{\bJ}
\right] \quad \Leftrightarrow \quad 
F_{\bG}  =\sum_{\bJ} C_{\bJ} -\omega_{\bG} + \frac{3}{2}  N_{\ext}
\;,
\eeq
where we restrict the study to the case $\bG\neq \emptyset$
so that $C_{\bJ}\geq 1$. But $C_{\bJ} = C_{\bG}$, then 
\beq
F_{\bG}  =3(C_{\bG}-1) -\omega_{\bG}+ 3 + \frac{3}{2}  N_{\ext}\;.
\label{facebor}
\eeq
On the other hand,  
consider the pinched jackets $\tJ$ of $\cexG$. 
From the fact that each face of the boundary graph 
$\bG$ (labeled by three colors, say $(0ab)$) is shared by exactly 
$2$ pinched jackets of the graph $\cexG$ (which will be labeled as 
$(0a\dots b)$ where the dots can be only the two remaining
numbers $\check a, \check b \in \{1,2,3,4\} \setminus \{a,b\}$; 
 see Fig.\ref{fig:boundface}),

we can relate 
\bea
\sum_{J} F_{\ext; \,\tJ} = 2F_{\bG}\;.
\label{fext}
\eea
Inserting (\ref{facebor}) into (\ref{fext}), then plugging the result
in (\ref{facint}), furthermore, noting that $ F_{\inter;\cG}  =F$
and $2L=6V_6 + 4V_4+2 (V_2  + V'_2 + V_2'') -N_{\ext}$, 
the divergence degree $\omega_d(\cG)$ (\ref{convergence}) 
can be recast in  the form 
\bea
\omega_d(\cG) &=&  -2 L + F_{\inter;\cG}  + 2 V_2'  \crcr
&=& 
- \frac13  [(3(C_{\bG}-1) -\omega_{\bG}+ 3 + \frac{3}{2}  N_{\ext})]
- \frac13 \sum_{J} g_{\tJ} 
+4 -V_4  -  2(V_2 + V_2'') \crcr
& =&  - \frac13 \sum_{J} g_{\tJ} +  \frac13  \sum_{\bJ} g_{\bJ} - (C_{\bG}-1) -V_4   -  2(V_2 + V_2'') - \frac{1}{2} (N_{\ext}  -6) 
\eea
which is the desired relation.
\qed 

The detailed analysis of the divergence degree 
is in order. An important part of that analysis is the 
understanding of the quantity
\bea
-\frac13\sum_{J} g_{\tJ} + \frac13\sum_{\bJ} g_{\bJ} - (C_{\bG}-1)\;.
\label{gen}
\eea
This is purpose of the next section. 

\section{Analysis of the Divergence Degree}
\label{sect:cdegan}

This section is divided in two 
parts: the first part addresses the study of the 
sign of the quantity (\ref{gen}) which is essential 
in the understanding of $\omega_d(\cG)$ and, 
based on this analysis, the second part classifies 
the primitively divergent graphs. 

\subsection{Bounds on genera}
\label{subsect:cdegan}

In this subsection, we restrict ourselves to the only important part of the problem, 
namely the analysis of graphs without  any two-point $V_2$, $V'_2$ and $V''_2$
vertices. Indeed, when such vertices are present, we can first contract any maximal chain
of these vertices into a single line, then analyze the resulting reduced graph, then 
reintroduce the chains and the full graph analysis follows easily.

We now introduce a new tool to study of the 
divergence degree, namely a sequence of contractions which generalizes the idea 
of dipole contraction
\cite{Gur3,Lins,FerriCag}. 
Contraction of dipoles separating bubbles in a colored theory
can be performed without changing the degree. In particular a tree
of 0-lines in a graph can always be contracted leading to a single 
``big'' bubble with many 0-loops attached, which is the tensor analog of a rosette graph.
This bubble is melonic if the initial vertices of the theory were melonic, which is the case here. 
In this subsection, we continue loop contractions on this generalized rosette
which may change the degree. Hence, we generalize to the tensor context the 
analysis by Filk moves \cite{Filk} of a ribbon rosette in non-commutative field theory \cite{Rivasseau:2007ab}. 

\begin{figure}
 \centering
     \begin{minipage}[t]{.7\textwidth}
      \centering
\includegraphics[angle=0, width=8cm, height=1cm]{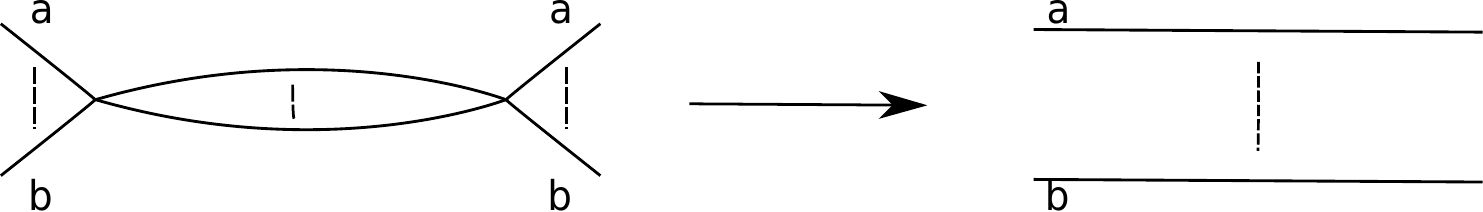}
\caption{ {\small $0k$-dipole contraction: external lines get glued. 
\label{cont}}}
\end{minipage}
\end{figure}

\begin{definition}[$0k$-dipole and contraction]
We define a $0k$-dipole, $k=0,1,\ldots,4$, as a maximal subgraph of 
$\cexG$ made of $k+1$ lines joining two vertices, 
one of which of color 0. Maximal means the  $0k$-dipole 
is not included in a $0(k+1)$-dipole. 

The contraction of a $0k$-dipole erases the $k+1$ lines 
of the dipole and joins the remaining $D-k$ lines on both 
sides of the dipole by respecting colors (see Fig.\ref{cont}).
\end{definition}

\begin{lemma}[Graph contraction]
\label{contract}
Performing  the maximal 
number $(6V_6 + 4 V_4 - N_{\ext})/2$ of $0k$-dipole contractions on $\cexG$ in any arbitrary
order and erasing the external legs of $\cG$ leads to the boundary graph $\bG$. 
\end{lemma}
\noindent{\bf Proof.} 
The graph $\cexG$ possesses $V_{\cexG}=6V_6 + 4 V_4$ vertices
which can be decomposed in internal vertices 
$V_{\inter;\cexG}= 6V_6 + 4 V_4 - N_{\ext}$ and external ones 
$N_{\ext}$. To each internal vertices corresponds 
one half-line with color $0$, therefore 
 $V_{\inter;\cG}/2 = L_{0;\inter;\cG}$ is the number of
lines of color $0$ on which the contraction procedure 
will be applied. 

Let us call the graph resulting from the contraction by $\widehat\cG$. 
Since all vertices and lines of $\bG$ are not concerned
by the procedure,\footnote{For instance, an open face $(0a)$ 
cannot be deleted by the dipole contraction procedure, it can just  be shortened.} they will appear again in $\widehat\cG$
and hence, obviously, $\bG\subset\widehat\cG$.
Furthermore, given any order of contraction, the final graph 
has exactly the same number of vertices and 
lines than $\bG$. Therefore, these graphs should coincide:
$\widehat\cG=\bG$.

\qed

We turn now to the proof of two local lemmas  which study the change in the sum over jackets of the difference in genera under a dipole contraction.
We consider a colored connected  graph $ \cexG$, a fixed 
$0k$-dipole and the contracted graph  $ \cexG' $, which may or may not be connected. We notice first that during the contraction the numbers of vertices and lines change as
\beq
V \to V' =V-2\;, \qquad
L \to L' = L-5 \;, 
\label{vv'}
\eeq
and the number of connected components can change from $c=1$ to    $ c' \leq 4  $.  Note however that
 $c'$ is constant for all jackets.  Indeed, 
all jackets have the same number of connected components
corresponding to the number of connected components 
of the graph obtained after contraction.

To track the change in faces, we introduce the notion of 
pair types for the contraction.

A pair for which the two colors are external to the dipole is called ``outer''. A pair 
which has one color inside the dipole and one out is called a ``mixed'' pair.
A pair with two colors inside the dipole is called an ``inner'' pair.
The total number of pairs is always 10 and the number of mixed pairs is at least 4.
A pair is said to belong to a jacket if the pair is one of the five adjacent pairs in the jacket cycle.

We say that an outer pair is of type A, or disconnected by the dipole contraction
if the half-strands at each corner on the left and on the right of the dipole belong to two different
connected components of the graph after the dipole contraction. 
In the converse case, we call it a ``special'' pair. A special 
pair can be single-faced if the two corners belong to the same face of the graph,
or double-faced if the two corners  belong to two different faces of the graph.
A moment of reflexion about open faces reveals that any type A outer pair 
must be single-faced at the beginning. Hence we have a classification
of outer pairs into three types:
\begin{itemize}
\item
Type A outer pairs are single-faced,

\item Type B outer pairs are single-faced,

\item Type C outer pairs are double-faced.
\end{itemize}

\begin{figure}
 \centering
     \begin{minipage}[t]{.7\textwidth}
      \centering
\includegraphics[angle=0, width=4cm, height=2cm]{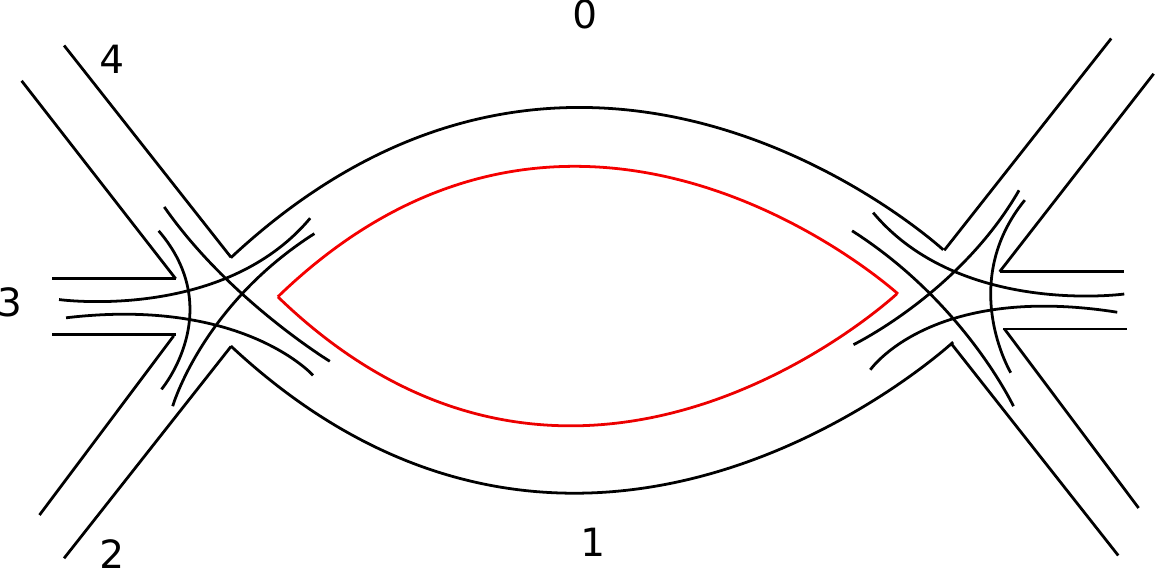}
\caption{ {\small An inner face (in red) of a $01$-dipole diagram.
\label{inface}}}
\end{minipage}
\end{figure}

Transverse pairs do not change their number of faces under contraction. Inner pairs have one face less
after contraction. Type A and B outer pairs have one face more after contraction and  type C outer pairs
have one face less  after the contraction.
Hence for any jacket
\beq
(F_{\tJ' } - F_{\tJ} )  =  \vert A_{\tJ} \vert + \vert B_{\tJ} \vert - \vert C_{\tJ} \vert - \vert I_{\tJ} \vert \;,
\label{ffprim}
\eeq
where $ \vert A_{\tJ} \vert $ is the number of pairs of type A in the jacket and so on, and $|I_{\tJ}|$ is the number of inner faces (for an illustration, see Fig.\ref{inface}).

We prove now two lemmas\footnote{ We warmly thank the referee
for his constructive remarks improving the formulation and 
proof of these lemmas.} analyzing the difference \eqref{ffprim} 
after summing over jackets.

\begin{lemma}
\label{localjacket} Performing any $0k$-dipole contraction
on a graph, we obtain 
\bea 
\sum_{J} (g_{\tJ} - g_{\tJ'})   \geq  0  \;.
\label{interm1}
\eea
Moreover, if  $\sum_{J}  (g_{\tJ} - g_{\tJ'})   >0$ then
\bea  \sum_{J}  (g_{\tJ} - g_{\tJ'})   \geq   6 \; .
\label{interm2}
\eea
\end{lemma}
\noindent{\bf Proof.}
 We have
\bea
2 - 2g_{\tJ} &=&  V - L+ F_{\tJ}\;, \crcr
2 c'- 2g_{\tJ' } &=&V' - L'+ F_{\tJ' } = (V-2) - (L-5) + F_{\tJ' }  \;,\crcr
(g_{\tJ} - g_{\tJ'})  & =& \frac{1}{2}[ (F_{\tJ' } - F_{\tJ} )  +3  - 2(c'-1)]\;.
\label{gmgprim}
\eea
Now using the fact that each face is shared by 6 jackets 
and that $c'$ is constant for all jackets, by summing the last expression \eqref{gmgprim}
over all jackets (recalling there are 12 of them) and using 
\eqref{ffprim}, we infer
\beq
\sum_J (g_{\tJ} - g_{\tJ'})  = 3(A+B-C-I)
  +18  - 12(c'-1)
\eeq
where we introduced the quantities 
$A = \sum_J A_{\tJ}$, $B = \sum_JB_{\tJ}$, $C=\sum_JC_{\tJ}$,
and $I=\sum_JI_{\tJ}$.

We perform a case by case study, proving that $  3(A+B-C-I)
  +18  - 12(c'-1)$  is always positive and, always greater than $6$
whenever it turns out to be strictly positive.

\medskip
\noindent{ $\bullet$ \bf 1rst Case:  00-dipole contraction.} An unique internal line with color $0$ is contracted. 
There are four mixed pairs, six outer pairs and no inner pair. Each jacket contains two mixed and three outer pairs.

\noindent
{\it - 1rst subcase $c'=4$ (Fig.7A)}.  
This can happen only if the resulting
graph has on each line $1,2,3,4,$ a connected two-point subgraph.
In that case, all six outer pairs must be of type $A$.  
Hence, the $00$-dipole contraction yields for all jackets:  
\beq
 3(6) +18  - 12(c'-1) =0\;.
\eeq

\noindent
{\it - 2nd subcase  $c'=3$ (Fig.7B)}. This case happens if we have two connected two-point 
functions plus one connected four-point function
on four half-lines hooked to the dipole. In that case, we have
5 corner pairs of type A and one special pair, which can be type B or type C. We symbolically write for any possible choices: 
\beq  \label{special1}
3(5 \pm 1)   +18  - 12(c'-1) \in \{6, 12\}\; .
\eeq

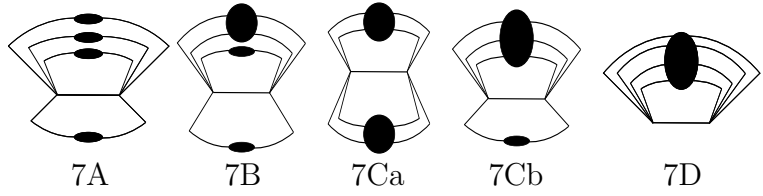
\begin{figure}
 \centering
     \begin{minipage}[t]{.7\textwidth}
      \centering
\includegraphics[angle=0, width=10cm, height=2cm]{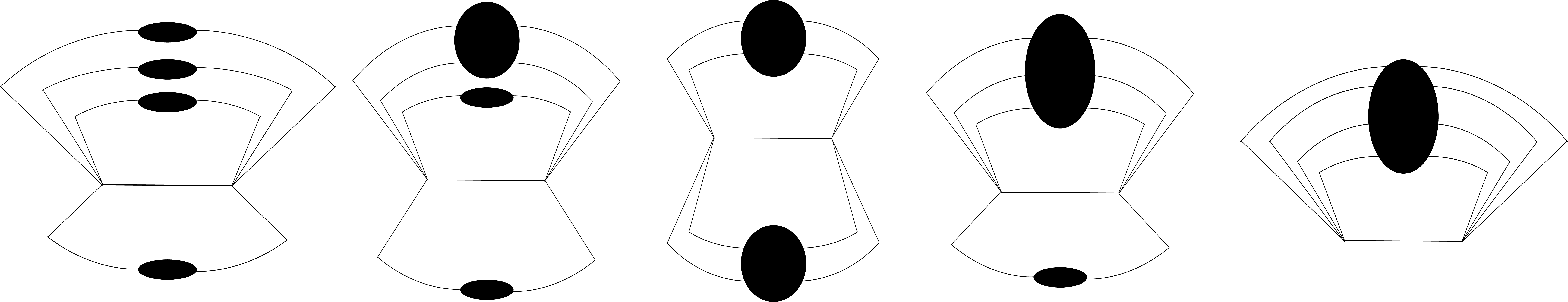}
\vspace{0.1cm}
\caption{ {\small $00$-dipoles configurations.}}
\end{minipage}
\put(-285,-12){7A}
\put(-227,-12){7B}
\put(-179,-12){7Ca}
\put(-127,-12){7Cb}
\put(-61,-12){7D}

\end{figure}

\noindent
{\it - 3rd subcase $c'=2$ (Fig.7Ca and 7Cb)}. This can happen with two
subsubcases: one with two connected four-point functions
and one with one six-point and one two-point connected functions.

In the first subsubcase, there are 4 corner pairs of type A,
and 2 of type either B or C. We have
\beq
 3(4\pm 1 \pm 1)   +18  - 12(c'-1) \in \{12,18,24\}\;.
\eeq

In the second subsubcase,  there are 3 outer pairs of type A and three special pairs. This corresponds to
\beq
 3(3\pm 1 \pm 1 \pm 1 )   +18  - 12(c'-1) \in \{6,12,18,24\}\;.
\eeq

\noindent{\it - 4th subcase  $c'=1$ (Fig.7D)}. Contracting the dipole gives a single 
connected component, hence $c'-c=0$. It can happen if 
we have a eight-point connected function. 
There are no pairs of type A and six special pairs, hence
\beq 
 3(\pm 1 \pm 1 \pm 1 \pm 1 \pm 1 \pm 1)   +18  - 12(c'-1) \in \{0,6,12,18,24,30,36\} \;.
\eeq

\medskip
\noindent{ $\bullet$ \bf 2nd Case:  01-dipole contraction.}
There is one inner pair, six mixed pairs and three outer pairs.
There are three subcases. 

\noindent{\it - 1rst subcase $c'=3$}. In this case, the three outer pairs are type A, and there are no special pairs:
\beq
  3(3-1)   +18  - 12(c'-1) = 0\;.
\eeq

\noindent{\it - 2nd subcase  $c'=2$}.
This situation yields two outer pairs of type A and one  special:
\beq
  3(2\pm 1 -1)   +18  - 12(c'-1) \in \{6,12\}\;.
\eeq

\noindent{\it - 3rd subcase  $c'=1$}.
Here, no outer pair is type A, the three outer pairs are special.
This yields
\beq
  3(\pm 1 \pm 1 \pm 1 - 1)  +18  - 12(c'-1)\in \{6,12,18,24\} \;.
\eeq

\medskip
\noindent{ $\bullet$ \bf  3rd Case:  02-dipole contraction.}
There are three inner pairs, six mixed pairs and one outer pair. 
There are two subcases. 

\noindent{\it  - 1rst subcase $c'=2$}. In this case, the outer pair is type A, and in all cases
\beq
 3(1-3) +18  - 12(c'-1) =0\;.
\eeq

\noindent{\it - 2nd subcase  $c'=1$}.
Here, the outer pair is special and one gets
\beq
3(\pm 1 -3) +18  - 12(c'-1) \in \{6,12\}\; .
\eeq

\medskip
\noindent{ $\bullet$ \bf  4th Case:  03-dipole contraction}.
There are six inner pairs and four mixed pairs, $ (c'-1)  = 0$
such that one has 
\beq
3(-6) +18  - 12(c'-1)  = 0\; .
\eeq

\medskip
\noindent{ $\bullet$ \bf  5th Case: 04-dipole contraction.}  This is the easiest case as
it destroys completely a full vacuum connected component
with two vertices and five lines.
In that case, there are ten inner pairs, $ (c'-1)  = -1$ and so
\beq
3(-10) +18  - 12(c'-1)  = 0 \;.
\eeq
Hence in all above cases \eqref{interm1} and \eqref{interm2} are true. 

\qed

\begin{definition}[Jacket inclusion]
We say that a $4$-jacket $J$ 
(i.e. a jacket defined by a cycle of length 5 up to orientation) 
contains a $3$-jacket $J'$ without the color $0$, and we write $J'\subset  J$, if $J'$
is the cycle obtained by contracting the color $0$ is the
cycle of $J$ (up to orientations). 
\end{definition}
There are obviously 4 jackets $J$, namely $(0abcd), (a0bcd), (ab0cd)$ and $  (abc0d)$  containing 
a given  (boundary jacket) $J' =(abcd)$. They correspond
to inserting $0$ at any position in $J'$. 

\begin{lemma}[Genus bounds]
\label{genbound} 
We have
\beq  \label{interm3}
\sum_{J} g_{\tJ} -  4  \sum_{\bJ} g_{\bJ} \in \N\; .
\eeq
Moreover
\bea \label{interm4}\sum_{\bJ} g_{\bJ} >0
\quad  &\Rightarrow &\quad  \sum_{J} g_{\tJ} -  4  \sum_{\bJ} g_{\bJ}  \ge 6\;,  \\
 \sum_{\bJ} g_{\bJ} =0\;\; {\rm and} \;\; \sum_{J} g_{\tJ} >0
\quad & \Rightarrow &\quad  \sum_{J} g_{\tJ}   \ge 6 \;.
\label{interm5}
\eea
\end{lemma}
 \noindent{\bf Proof.}   
We perform a full sequence of $0k$-dipole contractions on the initial graph $\cG$
and arrive at the graph $\widehat \cG = \bG$.
By Lemma \ref{localjacket}, any genus of any pinched jacket $\tJ$ decreases along that sequence
\footnote{ Given a jacket $\tJ$ 
and its contraction $\tJ'$, we can also prove that  $g_{\tJ}\geq g_{\tJ'}$
using similar techniques as developed in Lemma \ref{localjacket}. 
This inequality, holding jacket by jacket,
is a stronger result than Lemma \ref{localjacket}.}
 and, at the end, the pinched jacket coincides with the  
jacket $\bJ$. To each boundary jacket $\bJ$ we can 
associate four $\tJ$ such that $\bJ\subset \tJ$. 
This  proves  \eqref{interm3}-\eqref{interm4}. 
If $ \sum_{\bJ} g_{\bJ} = 0$ and $ \sum_{J} g_{\tJ} >0$, then at some point  along that sequence
we can again use \eqref{interm2}, 
which proves \eqref{interm5}.
\hfill $\square$

\subsection{Classification of divergent graphs}
\label{subsect:conv}

We have $\bG \neq \emptyset$, hence $C_{\bJ} \geq 1$,
this means that we will always consider a graph $\cG$ with a boundary in the following developments. 
Furthermore $C_{\bJ} \leq N_{\ext}/2$,
because each connected components must have at least 
a non zero even number of external legs.
Let us define the integer $P(\cG)= (C_{\bG}-1)  + V_4
+2(V_2+ V_2'') + \frac12  \left[ N_{\ext} - 6\right]$.
Lemma \ref{genbound} translates into
\begin{lemma}[Power counting bound]
\label{powcbound}
We have
\bea 
\label{betterbound1}
\omega_d(\cG) = -\frac13 \left[ 
\sum_{J} g_{\tJ} - \sum_{\bJ} g_{\bJ} \right]-P(\cG) &\leq&   -  \sum_{\bJ} g_{\bJ}   -P(\cG)\,  ,  \\
 \label{betterbound2}
\sum_{\bJ} g_{\bJ}   > 0 \quad  &\Rightarrow& \quad \omega_d(\cG)
\leq -2  - \sum_{\bJ} g_{\bJ} -P(\cG) \, ,  \\
 \sum_{\bJ} g_{\bJ} =0\;\; {\rm and} \;\; \sum_{J} g_{\tJ} >0 \quad & \Rightarrow& \quad \omega_d(\cG)
\leq -2  -P(\cG) \,   . 
\label{betterbound3}
\eea
\end{lemma}

We search now for the list of graphs with  $\omega_d(\cG)  \geq 0$ which are
those which should be renormalized.

\noindent{\bf Case} $N_{\ext} > 6$: 
In this situation, $N_{\ext}\geq 8$, so that $P(\cG) \geq 1$
$\omega(\cG) \leq -1$ and hence the graph
has a converging amplitude. 

\noindent{\bf Case}  $N_{\ext} = 6$:
The divergence degree is at most zero and can be so only if
\beq  C_{\bJ} = 1\;, \quad \sum_{\bJ} g_{\bJ} =\sum_{J} g_{\tJ} = 0\;,  \quad  V_4 =V_2+ V_2'' = 0\,.
\eeq

\noindent{\bf Case} $N_{\ext} = 4$: $P(\cG)= (C_{\bG}-1)  + V_4
+2(V_2+ V_2'') -1$. The divergence degree is at most 1. It can be 1 only if $P(\cG)=-1$, and in fact if 
\beq  C_{\bJ} = 1\;, \quad \sum_{\bJ} g_{\bJ} =\sum_{J} g_{\tJ} = 0\;,  \quad  V_4 =V_2+ V_2'' = 0\;.
\eeq
But it could be zero if $P(\cG)=0$, in which case we must have either
\beq  C_{\bJ} = 2\;, \quad \sum_{\bJ} g_{\bJ} =\sum_{J} g_{\tJ} = 0\;,  \quad  V_4 =V_2+ V_2'' = 0\;,
\eeq
or
\beq  C_{\bJ} = 1\;, \quad \sum_{\bJ} g_{\bJ} =\sum_{J} g_{\tJ} = 0\;,  \quad  V_4 =1\;,  \quad V_2+ V_2'' = 0\;.
\eeq
Finally, when $P(\cG)=-1$, hence $C_{\bJ} = 1$, $V_4 =V_2+ V_2'' = 0$, if $\sum_{\bJ} g_{\bJ}  >  0$,
we have $\omega_d(\cG) \leq -2 $ by  \eqref{betterbound2} and if
$ \sum_{\bJ} g_{\bJ} =0\;\; {\rm and} \;\; \sum_{J} g_{\tJ} >0$ we have
$\omega_d(\cG) \leq -1 $ by  \eqref{betterbound3}.

\noindent{\bf Case} $N_{\ext} = 2$: $P(\cG)= (C_{\bG}-1)  + V_4
+2(V_2+ V_2'') -2$. In that case $\sum_{\bJ} g_{\bJ}  =  0$ since the only possible 
colored boundary graphs made with two external vertices is the standard one with 
six planar jackets. The analysis is slightly lengthy and we get 5 possible cases 
of divergent graphs.

In summary, the divergent graphs are determined by the following
table:

\begin{center}
\begin{tabular}{lccccc||cc|}
$N_{\ext}$ & $V_2 + V_2''$ & $V_4$ & $\sum_{\bJ} g_{\bJ}$ & $C_{\bG}-1$ & $\sum_{\tJ}  g_{\tJ}$ & $\omega_d(\cG)$  \\
\hline\hline
6 &0 & 0 & 0 & 0& 0& 0\\
\hline
4 & 0 & 0 & 0 & 0 & 0 & 1 \\
4 & 0  & 1 & 0 & 0 & 0 & 0\\
4 & 0 & 0 & 0 & 1 & 0 & 0 \\
\hline
2 & 0 & 0 & 0 & 0 & 0 & 2\\
2 & 0 & 1 & 0 & 0 & 0 & 1\\
2 & 0 & 2 & 0 & 0 & 0 & 0\\
2 & 0 & 0 & 0 & 0 & 6 & 0\\
2 & 1 & 0 & 0 & 0 & 0 & 0\\
\hline\hline
\end{tabular}

\vspace{0.2cm}
Table 1
\end{center}

\subsection{The $\int\varphi^2\int\varphi^2 $ anomalous term}
\label{sect:phi2phi2}

\begin{figure}
 \centering
     \begin{minipage}[t]{.8\textwidth}
      \centering
\includegraphics[angle=0, width=6cm, height=4.5cm]{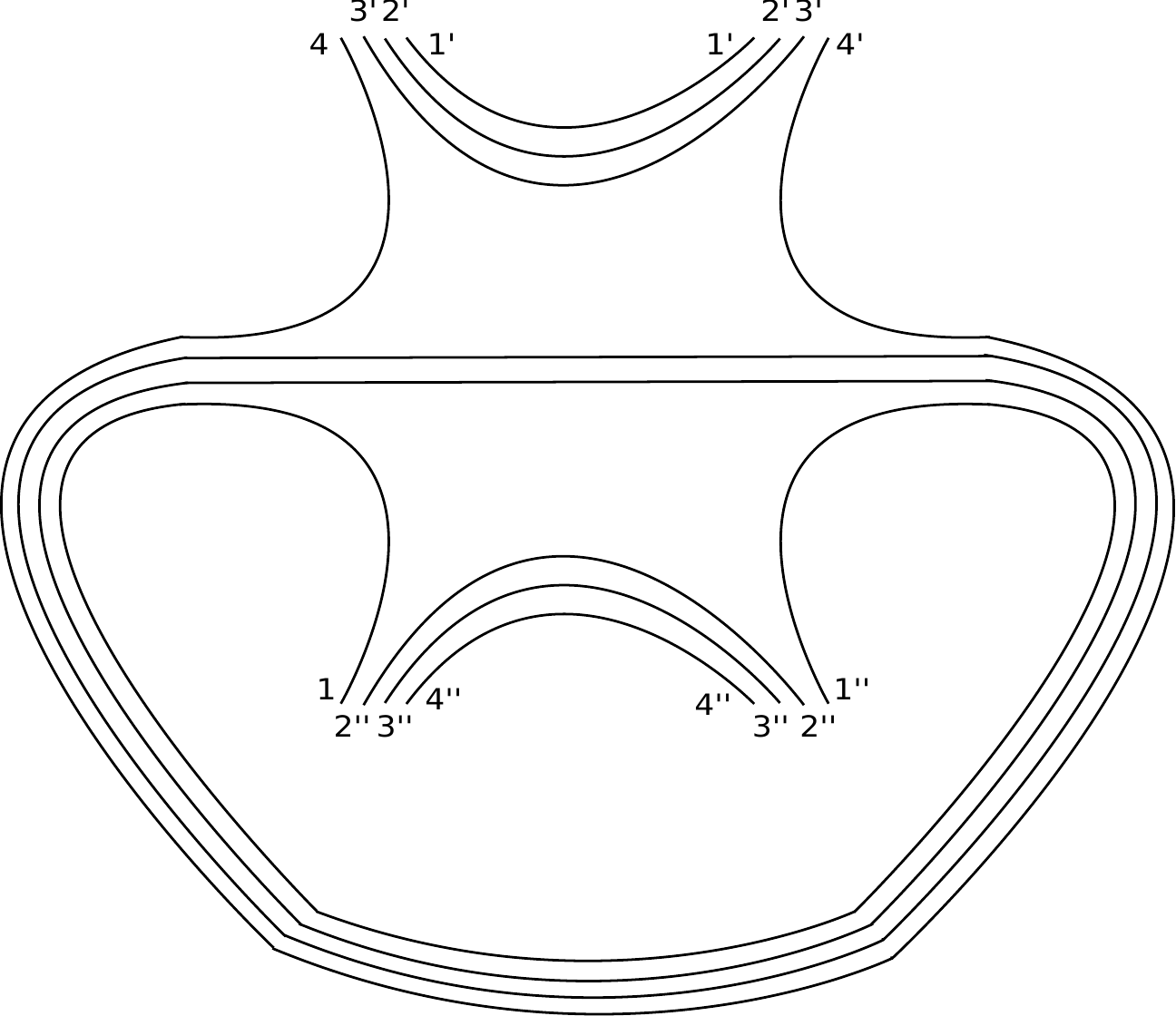}
\caption{ {\small The tadpole of $V_{6;2}$ has a disconnected
boundary graph.
\label{anom}}}
\end{minipage}
\end{figure}

Since $V''_2$ is even, no graph with internal counterterm of the $V''_2$
can appear in the previous table. 
However the fourth row of the table corresponds to melonic graphs with disconnected
boundary graphs, they do appear and are really divergent. The first and lowest order example
example is the special ``diagonal tadpole" built on the $V_{6;2}$ vertex (see Fig.\ref{anom}). This graph has one line and two internal faces $0a$ and $0b$ where $a$ and $b$ are the 
colors of the two ``inner strands" in the $V_{6;2}$ interaction. Hence $-2L + F =0$. The graph is really 
logarithmically divergent as the propagator $1/ (p^2 + m^2)$ is positive, hence there is no way any 
unexpected cancellation could affect its amplitude.

It is difficult to interpret yet this anomalous term but it happens in 4 dimensions and not in 3. 
This kind of factorized log-divergent  term is best represented as an integral over an \emph{intermediate field} as
\beq
 e^{- ( \int \varphi^2)^2} = 
c\int d\sigma \; e^{ -\int \sigma^2  - 2i \int \sigma \varphi^2}\;,
\eeq
$c$ being some constant, and whose propagator is the dotted line in 
Fig.3. It joins the two two-point functions. 
But this propagator does not have any strand, hence our gravity theory
generates a scalar matter field.

\section{Renormalization}
\label{sect:renorm}

We now implement the renormalization program for the $p$-point functions
which are divergent and characterized as given by Table 1. We use 
Taylor expansions around the local parts in direct space 
in the manner of \cite{Rivasseau:1991ub,Gurau:2005gd}.

\subsection{Renormalization of the six-point function}
\label{subsect:sixpt}

Consider a general six-point function subgraph $G^{(k)}_i$, 
namely with $N_{\ext}(G^{(k)}_i)=6$ of the type of 
the first line  of Table 1. Since $\sum_{\bJ} g_{\bJ}=0$,
we know that the boundary graph is itself a melonic 
graph and hence the pattern of external positions 
follows the form either of $V_{6;1}$ or of $V_{6;2}$. 

We reintroduce the graph with external propagators 
and call its amplitude $\bar A_6(G^{(k)}_i)$. 
External positions variables are labeled by $\theta^{\ext}_{l,s}$,  $l=1,2,3,4,5,6$ and $s=1,2,3,4$,
where $s$ is, as usually, the strand index while $l$ can be considered
as the external leg index with scale $j_l$. Recall that $j_l$ indices are 
strictly smaller than $i$ the index of $G^{(k)}_i$. 
 $\theta^0_{l,s}$ denotes the position connected 
to the external  end-point $\theta^\ext_{l,s}$.

The following procedure is standard  \cite{Rivasseau:1991ub} and consists
in performing a Taylor expansion in direct space by interpolating
moves of the external legs. Remark that this interpolation
should be periodic and consistent with the fact that we are dealing with
a torus. For convenience, we also change the local parametrization and integration bounds of the Haar measure to be $[-\pi,\pi)$.
Hence, we interpolate $\theta^{0}_{l(v),s}$ using a
parameter $t \in [0,1]$ such that, for $\theta^{0}_{ l'(v'), s}\in [0,\pi)$
\bea
\theta^0_{l(v),s} \in [\theta^{0}_{ l'(v'), s}-\pi, \pi)\,,&&
\theta^0_{l(v),s}= \theta^{0}_{ l'(v'), s}  
+ t (\theta^0_{l(v),s}  -\theta^{0}_{l'(v'), s}) \Big{|}_{t=1} \;, 
\crcr
\theta^0_{l(v),s} \in [-\pi,\theta^{0}_{ l'(v'), s}-\pi)\,,&&
\theta^0_{l(v),s}= \theta^{0}_{ l'(v'), s} -2\pi
+ t (\theta^0_{l(v),s}  -\theta^{0}_{l'(v'), s} +2\pi) \Big{|}_{t=1} \;,
\label{interpol}
\eea
where $\theta^{0}_{ l'(v'), s}$ is the internal position connected 
to the external index $\theta^{\ext}_{l', s}$ which can be associated
with $\theta^\ext_{l,s} $ according to the particular pattern of  the $\varphi^6$ vertices. From the above, one can easily infer
the interpolation for the other range of values $\theta^{0}_{ l'(v'), s}\in [-\pi,0)$. 

We relate the lines $l$ and $l'$ and their strand
index $s$ in two possible ways dictated by the boundary 
graphs of a $\varphi^6$ form:

(1) the couples $(\theta^0_{l,s},\theta^{0}_{l', s})$ for 
$(l,l')\in\{(2,1),(4,3),(6,5)\}$ are connected 
with respect to the strand indices $s=2,3,4$, whereas pairs 
$(l,l')\in\{(1,6),(5,4),(3,2)\}$ will be connected only for the strand index $s=1$; 
performing a permutation on the role of $s=1,2,3$ and $4$
gives the parametrization for remainder vertices of $V_{6;1}$;  

(2) the couples $(\theta^0_{l,s},\theta^{0}_{ l', s})$ for  $(l,l')\in\{(3,2),(6,5)\}$ are connected 
for strand indices $s=2,3,4$, $(l,l')=(4,1)$ connected 
for $s=2,3$, $(l,l')\in\{(2,1),(4,3)\}$ are connected for a single
index $s=4$ and $(l,l')\in\{(5,4),(1,6)\}$ 
are connected for $s=1$; 
a permutation on the role of $(2,3)$ 
for any other couple in $\{1,2,3,4\}$ yields
the parameterizations for remainder vertices as defined
by $V_{6;2}$.

In the following, we will focus on the vertex of the first kind defined
by pairings (1). For the second kind, it can be checked that similar results will be also valid. 

Consider the amplitude of the subgraph $G^{(k)}_i$ with 
external propagators characterized  as above given by
(in simplified notations)
\bea
\bar A_6(G^{(k)}_i)[\{ \theta^\ext_{l,s}\}  ] &=& 
 \int \left\{ [ \prod_{\ell} d\theta_{\ell, s}]\, \Big[\prod_{l} 
C_{j_l}(\{\theta^\ext_{l, s}\};\{ \theta^0_{l(v),s} \})\Big] \right. \crcr
&&\left. \Big[   \prod_{\ell \neq l} C_{i_\ell}(\{\theta_{\ell(v), s}\};\{ \theta_{\ell(v'), s} \})\Big] 
\prod_{v\in \cV}  
\delta(\theta_{v, s} - \theta_{v ,s'}) \right\}.
\label{amplst}
\eea 
We introduce a function $\bar A_6(G^{(k)}_i)[\{ \theta^\ext_{l,s}\};t] $ 
and write the amplitude (\ref{amplst})  as 
$
\bar A_6(G^{(k)}_i)[\{ \theta^\ext_{l,s}\};$ $ t=1  ]= \bar A_6(G^{(k)}_i)[\{ \theta^\ext_{l,s}\}; t =0 ]  + \int_0^{1} dt\;
\frac{d}{dt} \bar A_6(G^{(k)}_i)[\{ \theta^\ext_{l,s}\}; t  ],
$
where the function $\bar A_6(G^{(k)}_i)[\{ \theta^\ext_{l,s}\};t] $
is obtained by interpolating (\ref{amplst}) using (\ref{interpol}), namely
\bea
\bar A_6(G^{(k)}_i)[\{ \theta^\ext_{l,s}\};t] &= &
 \int [ \prod_{\ell} d\theta_{\ell, s}] \; \Big\{\Big[ \prod_{l=1}^6 C_{j_l}(\{\theta^\ext_{l, s}\};\{ \theta^{0}_{ l'(v'), s}  
+ t (\theta^0_{l(v),s}  -\theta^{0}_{l'(v'), s})\})\Big]   \,\crcr
&& 
\Big[ \prod_{\ell \neq l } 
C_{i_\ell}(\{\theta_{\ell(v), s}\};\{ \theta_{\ell(v'), s}\})\Big]
\prod_{v\in \cV}  
\delta(\theta_{v, s}-\theta_{v, s'}) \Big\} . 
\label{interpolam}
\eea
Henceforth, $\{\theta^{0}_{ l'(v'), s}  
+ t (\theta^0_{l(v),s}  -\theta^{0}_{l'(v'), s})\}$ denotes
any formula of the periodic interpolation (\ref{interpol}) according to the range of values of the coordinates. 

The term at $t=0$ is given by
\bea
&& \bar A_6(G^{(k)}_i)[\{ \theta^\ext_{l,s}\};0]=   \int \Big\{ [ \prod_{\ell \neq l} d\theta_{\ell, s}]\,
\Big[ 
\prod_{l=1}^6 C_{j_l}(\{\theta^\ext_{l, s}\};\{ \theta^{0}_{ l'(v'), s}\})\Big] 
\crcr
&&\int \Big[
[\prod_{l=2p+1} d\theta^0_{l,1}]
[\prod_{l=2p} d\theta^0_{l,2}d\theta^0_{l,3}d\theta^0_{l,4}]\Big] 
\Big[ \prod_{\ell \neq l } 
C_{i_\ell}(\{\theta_{\ell(v), s}\};\{ \theta_{\ell(v'), s}\})\Big] \prod_{v\in \cV}  
\delta(\theta_{v, s}-\theta_{v, s'}) \Big\}\crcr
&&
\prod_{l=1}^6 C_{j_l}(\{\theta^\ext_{l, s}\};\{ \theta^{0}_{ l'(v'), s}\}) =\crcr
&&
 C_{j_1}(\{\theta^\ext_{1,s}\};\{\theta^{0}_{6, 1}, \theta^{0}_{1,2} ,\theta^{0}_{1,3}, \theta^{0}_{1,4}\}) 
C_{j_2}(\{\theta^\ext_{2,s}\};\{\theta^{0}_{2, 1}, \theta^{0}_{1,2} ,\theta^{0}_{1, 3}, \theta^{0}_{1,4}\}) \crcr
&&
C_{j_3}(\{\theta^\ext_{3,s}\};\{\theta^{0}_{2,1}, \theta^{0}_{3,2} ,\theta^{0}_{3, 3}, \theta^{0}_{3,4}\}) 
C_{j_4}(\{\theta^\ext_{4,s}\};\{\theta^{0}_{4, 1}, \theta^{0}_{3,2} ,\theta^{0}_{3, 3}, \theta^{0}_{3,4}\})  \crcr
&&
C_{j_5}(\{\theta^\ext_{5,s}\};\{\theta^{0}_{4,1}, \theta^{0}_{5,2} ,\theta^{0}_{5, 3}, \theta^{0}_{5,4}\})
C_{j_6}(\{\theta^\ext_{6,s}\};\{\theta^{0}_{6,1}, \theta^{0}_{5,2} ,\theta^{0}_{6, 3}, \theta^{0}_{6,4}\}) \;,
\label{at0}
\eea
where we remove the possible $2\pi$ coming from the interpolation 
using the periodicity of the functions. 
Next, the remainder finds the following expansion
\bea
R_6
&=&
  \int_0^1 dt  \int \Big\{ [ \prod_{\ell} d\theta_{\ell, s}]
\Big( \sum_{l=1}^6 [\prod_{p\neq l} 
C_{j_{p}} (\{\theta^\ext_{p, s}\};\{ \theta^{0}_{ p'(v'), s}  
+ t (\theta^0_{p(v),s}  -\theta^{0}_{p'(v'), s})\})] \crcr
&& T_l \triangleright  C_{j_{l}} (\{\theta^\ext_{l s}\};\{ \theta^{0}_{ l'(v'), s}  
+ t (\theta^0_{l(v),s}  -\theta^{0}_{l'(v'), s})\})\,\Big)\crcr
&&\Big[ \prod_{\ell \neq l } 
C_{i_\ell}(\{\theta_{\ell(v), s}\};\{ \theta_{\ell(v'), s}\})\Big]\prod_{v\in \cV}  
\delta(\theta_{v, s}-\theta_{v, s'}) \Big\},
\label{rmaind}
\eea
where the operator $T_l$ differentiates with respect to particular
strands according to the vertex pattern and is given by
\beq
T_{l} = \sum_{k=0}^3\Big[ 
\delta_{l,2k+1} (\theta^0_{l(v),1}  -\theta^{0}_{l'(v'), 1} + r)
\partial_{\theta;1} 
+
\delta_{l,2k} \sum_{s=2}^4 
(\theta^0_{l(v),s}  -\theta^{0}_{l'(v'), s} + r)
\partial_{\theta;s} \Big],
\label{top}
\eeq
where $\partial_{\theta;s}$ is a partial derivative with
respect to the second set of arguments of $C_{j_l}$ 
containing $\theta^0$'s and taken at the strand $s$
and $r=\pm 2\pi$ or $0$ according to the sector of interpolation.
It remains to analyze these terms.  
Considering (\ref{at0}), the following statement holds
\begin{lemma}
\label{lemfacto6}
The quantity 
\beq
\int \Big[
[\prod_{l=2p+1} d\theta^0_{l,1}]
[\prod_{l=2p} d\theta^0_{l,2}d\theta^0_{l,3}d\theta^0_{l,4}]\Big] 
\Big[ \prod_{\ell \neq l } 
C_{i_\ell}(\{\theta_{\ell(v), s}\};\{ \theta_{\ell(v'), s}\})\Big] \prod_{v\in \cV}  
\delta(\theta_{v, s}-\theta_{v, s'})
\eeq
does not depends on $\{\theta^0_{l',s}\}$ defined by 
$\prod_{l=1}^6 C_{i_l}(\{\theta^{\ext}_{l, s}\};\{ \theta^{0}_{ l'(v'), s}\}) $.
\end{lemma}
\noindent{\bf Proof.} This is a consequence of translation invariance
of the propagators in spatial coordinates that we now review quickly. Having performed
a Taylor expansion of the interpolated amplitude (\ref{interpolam}), 
the zeroth order term is of the form (\ref{at0}), where  
the set arguments $\{\theta^{0}_{ l'(v'), s}\}$ present in the product $\prod_{l=1}^6 C_{i_l}(\{\theta^{\ext}_{l, s}\};\{ \theta^{0}_{ l'(v'), s}\})$ may be still involved in the internal structure. 
For simplicity, we focus on $\theta^0_{6,1}$ 
and we can consider the external face formed by successive positions
$\theta^0_{1,1}, \theta_{l_1,s_1}, \theta_{l_2,s_2}, \dots,\theta_{l_q,s_q}, \theta^0_{6,1}$. The propagators generating the face amplitude
associated with this sequence are functions of the differences $(\theta_{l_{\alpha},s_\alpha} - \theta_{l_\beta,s_\beta})$. 
Since $\theta^0_{1,1}$ and $\theta^0_{6,1}$ are external end-points,
we can always perform a change of variable $\widetilde \theta_{l_\beta,s_\beta} = \theta_{l_\alpha,s_\alpha} - \theta_{l_\beta,s_\beta}$ to remove
one of these external position labels. 
Note that, since we are dealing with a compact space, 
the bounds  of integration of the new variables $\widetilde\theta_{l_\alpha,s_\alpha}$ change. Nonetheless, recall that the propagators here
are periodic so that all these integration bounds can be translated indifferently 
to $[-\pi,\pi)$. 
In the present situation, choosing to remove $\theta^0_{1,1}$, we obtain
a face amplitude independent of that variable. Reproducing the argument
for each external faces, one proves the lemma.  

\qed

\begin{lemma}
\label{lemrem6}
The remainder $R_6$ of the amplitude interpolation 
can be bounded by
\beq
|R_6| \leq K M^{-( i(G^{(k)}_i) - e(G^{(k)}_i) )}  M^{\omega(G^{(k)}_i)} \;,\qquad
e(G^{(k)}_i)  = \sup_{l \, \text{external to}\; G^{(k)}_i } j_l \;,\qquad
 i(G^{(k)}_i) = \inf_{l \in G^{(k)}_i} i_l\;,
\eeq 
for some constant $K$.
\end{lemma}
\noindent{\bf Proof.} 
Let us first make a remark
concerning the integration bounds due to the splitting introduced
by the interpolation which is in the rough form (focusing on $r=+2\pi$)
\beq
 \int_{0}^{\pi} d\theta^0_{l',s} \int^{\pi}_{\theta^0_{l',s}-\pi} d\theta^0_{l,s} 
(\theta^0_{l(v),s}  -\theta^{0}_{l'(v'), s} ) \prod C 
 + \int_{0}^{\pi} d\theta^0_{l',s} 
\int_{-\pi}^{\theta^0_{l',s}-\pi} d\theta^0_{l,s} 
(\theta^0_{l(v),s}  -\theta^{0}_{l'(v'), s} + 2\pi ) \prod C \;.
\eeq
We can perform a change of variable in the term 
$\hat \theta^0_{l(v),s}  = \theta^0_{l(v),s} + 2\pi$ for which, 
clearly, the products of covariances and delta functions remain
invariant such that 
\beq
 \int_{0}^{\pi} d\theta^0_{l',s} \int^{\pi}_{\theta^0_{l',s}-\pi} d\theta^0_{l,s} 
(\theta^0_{l(v),s}  -\theta^{0}_{l'(v'), s} ) \prod C 
 + \int_{0}^{\pi} d\theta^0_{l',s} 
\int_{\pi}^{\theta^0_{l',s}+\pi} d\hat\theta^0_{l,s} 
(\hat\theta^0_{l(v),s}  -\theta^{0}_{l'(v'), s} ) \prod C\;. 
\eeq
By summing the two internal integrals we get a single 
integral as $\int^{\theta'+ \pi}_{\theta'-\pi} d\theta^0_{l,s} (\theta^0_{l(v),s}  -\theta^{0}_{l'(v'), s} )$. What we have gained here
is that the final integral can be fully bounded in terms of the difference
 $(\theta^0_{l(v),s}  -\theta^{0}_{l'(v'), s} )$.

Lemma \ref{lem1} yields a bound on the first derivative of the propagator (\ref{boundder}) as
\bea
\partial_{\theta,s} C_{j_{l}} (\{\theta^\ext_{s'}\};\{ \Theta_{s'} \})\, 
\leq K M^{3j_l} 
e^{-\delta M^{j_l}\sum_{s'}|\theta^\ext_{s'} -  \Theta_{s'} |} \;,
\eea
so that, taking the best estimate between external scales, 
 the following bound holds
\bea
|R_6| &\leq & K' M^{e(G^{(k)}_i)} M^{-i(G^{(k)}_i)}
  \int_0^1 dt  \int \Big\{ [ \prod_{\ell} d\theta_{\ell, s}]
\Big( \sum_{l=1}^6 [\prod_{p\neq l} M^{-2j_l}
e^{-\delta M^{j_p} \sum_s |\theta^\ext_{p, s}
- \theta^{0}_{ p'(v'), s}|}
] \crcr
&&
M^{2e(G^{(k)}_i)} e^{-\delta M^{j_l}\sum_{s'}|\theta^\ext_{l, s}- 
\theta^{0}_{ l'(v'), s}   |} \Big)
\Big[ \prod_{\ell \neq l } 
C_{i_\ell}(\{\theta_{\ell(v), s}\};\{ \theta_{\ell(v'), s}\})\Big]\prod_{v\in \cV}  
\delta(\theta_{v, s}-\theta_{v, s'}) \Big\}\,, \crcr
&&
\eea
we have used the facts that high internal decays entail 
$|(\theta^0_{l(v),s}  -\theta^{0}_{l'(v'), s})| \sim M^{-i_l}$
and hence, $|\theta^\ext_{p, s}- (\theta^{0}_{ p'(v'), s}  
+ t (\theta^0_{p(v),s}  -\theta^{0}_{p'(v'), s}))| \sim 
|\theta^\ext_{p, s} - \theta^{0}_{ p'(v'), s}|$ and, also,
that the distance
between the two internal positions, say $\theta^0_{l(v),s} $ and $\theta^0_{l'(v'),s}$,  can be optimized by choosing
$|\theta^0_{l(v),s} - \theta^0_{l'(v'),s}|\leq  M^{-i(G^{(k)}_i)}$.
As an effect, the integral in $t$ factors  and we get the result.

\qed

In conclusion, we have found that the 
zeroth order counterterm is given by (using Lemma \ref{lemfacto6}) 
\beq
\bar A_6(G^{(k)}_i)[\{ \theta^\ext_{l,s}\};0] =  \log M \int [\prod_{l'} d\theta_{l', s}]\,\prod_{l=1}^6 C_{j_l}(\{\theta^\ext_{l, s}\};\{ \theta^{0}_{ l', s}\})\;,
\eeq
hence is of the form  vertex $V_{6;1}$\footnote{ In fact, this 
a vertex $V_{6;1}$ with six external propagator
integrated to it.  } and is logarithmically divergent, whereas the sub-leading term is actually convergent due the power counting improvement by
$M^{-(i(G^{(k)}_i)- e(G^{(k)}_i))}$. This is exactly what is needed in order to 
perform the sum over the momentum assignments. 
One can easily check that performing
the similar analysis to other kind of permuted vertices $V_{6;1}$ or 
$V_{6;2}$ will lead to the same result.

\subsection{Renormalization of the four-point function}
\label{subsect:fourpt}

We use the same procedure as above in order 
to find the counterterms  of the four-point function
and for that consider a four-point function subgraph $G^{(k)}_i$, 
characterized by the one of the three lines  
of Table 1. Three cases may occur but, in all situations,
the graph (which should be melonic with a melonic boundary graph
in all cases)
has an external structure  either of the form $V_{4;1}$ or of the form 
$V_{4;2}$. The latter class includes graphs with disconnected
boundary graph (the last line of the table for $N_{\ext}=4$).

Let us call $\bar A_4(G^{(k)}_i)$ the amplitude associated
with $G^{(k)}_i$ equipped with external propagators.  
External position variables are labeled by $\theta^{\ext}_{l,s}$,  $l=1,2,3,4$ and $s=1,2,3,4$ and
external legs are at scale $j_l$.
We keep the same meaning of  $\theta^0_{l,s}$ as 
the positions connected to the external end-points $\theta^\ext_{l,s}$.

Interpolating $\theta^{0}_{l(v),s}$ using (\ref{interpol}),
 according to the particular pattern of external positions
of the boundary graph of the $\varphi^4$ type, 
we have:

(1) the couples $(\theta^0_{l,s},\theta^{0}_{l', s})$ for 
$(l,l')\in\{(2,1),(4,3)\}$ are connected 
with respect to the strand indices $s=2,3,4$, whereas pairs 
$(l,l')\in\{(1,4),(3,2)\}$ will be connected only for the strand index $s=1$; 
performing a permutation on the role of $s=1,2,3$ and $4$
gives the parametrization for remainder vertices of $V_{4;1}$;  

(2) the couples $(\theta^0_{l,s},\theta^{0}_{ l', s})$ for  $(l,l')\in\{(1,2),(3,4)\}$ are connected for all strand indices $s=1,2,3,4$ and this
defines the pattern of $V_{6;2}$.

Once again, we will only focus on the vertex of the first kind (1)
since the same reasoning will be valid for any other cases. 
The amplitude $\bar A_4(G^{(k)}_i)[\{ \theta^\ext_{l,s}\}]$ of the subgraph $G^{(k)}_i$ with external propagators (with above characteristics) is given by
a formula similar to (\ref{amplst}), and using external leg
interpolations giving the parametrized amplitude $\bar A_4(G^{(k)}_i)[\{ \theta^\ext_{l,s}\};t]$, 
we write
\bea
&&
\bar A_4(G^{(k)}_i)[\{ \theta^\ext_{l,s}\};t=1] =\\
&& \bar A_4(G^{(k)}_i)[\{ \theta^\ext_{l,s}\};t=0] 
+ \frac{d}{dt}\bar A_4(G^{(k)}_i)[\{ \theta^\ext_{l,s}\};t=0] 
 + \int_{0}^1 dt \;(1-t) \frac{d^2}{dt^2} \bar A_4(G^{(k)}_i)[\{ \theta^\ext_{l,s}\};t] \;,
\nonumber
\eea 
where the function $\bar A_4(G^{(k)}_i)[\{ \theta^\ext_{l,s}\};t] $ is given by
a quantity analog to (\ref{interpolam}), with four external 
propagators.
At $t=0$, we get the contribution
\bea
&& \bar A_4(G^{(k)}_i)[\{ \theta^\ext_{l,s}\};0]=   \int \Big\{ [ \prod_{\ell \neq l} d\theta_{\ell, s}]\,
\Big[ 
\prod_{l=1}^4 C_{j_l}(\{\theta^\ext_{l, s}\};\{ \theta^{0}_{ l'(v'), s}\})\Big] 
\crcr
&&\int \Big[
[\prod_{l=2p+1} d\theta^0_{l,1}]
[\prod_{l=2p} d\theta^0_{l,2}d\theta^0_{l,3}d\theta^0_{l,4}]\Big] 
\Big[ \prod_{\ell \neq l } 
C_{i_\ell}(\{\theta_{\ell(v), s}\};\{ \theta_{\ell(v'), s}\})\Big] \prod_{v\in \cV}  
\delta(\theta_{v, s}-\theta_{v, s'})\Big\}\;;\crcr
&&
\prod_{l=1}^4 C_{j_l}(\{\theta^\ext_{l s}\};\{ \theta^{0}_{ l'(v'), s}\}) = \crcr
&&
 C_{j_1}(\{\theta^\ext_{1,s}\};\{\theta^{0}_{4, 1}, \theta^{0}_{1,2} ,\theta^{0}_{1,3}, \theta^{0}_{1,4}\})
C_{j_2}(\{\theta^\ext_{2,s}\};\{\theta^{0}_{2, 1}, \theta^{0}_{1,2} ,\theta^{0}_{1, 3}, \theta^{0}_{1,4}\}) \crcr
&&
C_{j_3}(\{\theta^\ext_{3,s}\};\{\theta^{0}_{2,1}, \theta^{0}_{3,2} ,\theta^{0}_{3, 3}, \theta^{0}_{3,4}\}) 
C_{j_4}(\{\theta^\ext_{4,s}\};\{\theta^{0}_{4, 1}, \theta^{0}_{3,2} ,\theta^{0}_{3, 3}, \theta^{0}_{3,4}\})  \;.
\label{at04}
\eea
The second term is given by
\bea
&&
 \frac{d}{dt}\bar A_4(G^{(k)}_i)[\{ \theta^\ext_{l,s}\};t=0] 
=
  \int \Big\{ [ \prod_{\ell} d\theta_{\ell, s}]\;\;
\Big( \sum_{l=1}^6 [\prod_{p\neq l} 
C_{j_{p}} (\{\theta^\ext_{p, s}\};\{ \theta^{0}_{ p'(v'), s}\})] \crcr
&& T_l \triangleright  C_{j_{l}} (\{\theta^\ext_{l, s}\};\{ \theta^{0}_{ l'(v'), s}  \})\,\Big)
\Big[ \prod_{\ell \neq l } 
C_{i_\ell}(\{\theta_{\ell(v), s}\};\{ \theta_{\ell(v'), s}\})\Big]\prod_{v\in \cV}  
\delta(\theta_{v, s}-\theta_{v, s'}) \Big\} ,
\label{rmaind4}
\eea
where the operator $T_l$ now refers to 
\bea
T_{l} = \sum_{k=0}^2\Big[ 
\delta_{l,2k+1} (\theta^0_{l(v),1}  -\theta^{0}_{l'(v'), 1}+r)
\partial_{\theta;1} 
+
\delta_{l,2k} \sum_{s=2}^4 
(\theta^0_{l(v),s}  -\theta^{0}_{l'(v'), s}+r)
\partial_{\theta;s} \Big],
\eea
with $\partial_{\theta;s}$ and $r$ keeping their sense as in (\ref{top}). 
Finally and in the same anterior notations, the remainder computes to 
\bea
&&
R_4 = 
 \int_{0}^1 dt \;(1-t) 
 \int [ \prod_{\ell} d\theta_{\ell, s}] \; \Big\{ \crcr
&&\Big(
\sum_{l=1}^6 \Big[ 
\sum_{l'\neq l} [\prod_{q\neq l'} 
C_{j_{q}} (\{\theta^\ext_{q, s}\};\{ \theta^{0}_{ q'(v'), s} + t(\theta_{q(v),s} - \theta_{q'(v'),s} )\}) ] \crcr
&&
\times 
T_{l'} \triangleright C_{j_{l'}}(\{\theta^\ext_{l' s}\};\{ \theta^{0}_{ l''(v'), s}  
+ t (\theta^0_{l'(v),s}  -\theta^{0}_{l''(v'), s})\}) \crcr
&&\times
T_l \triangleright C_{j_l}(\{\theta^\ext_{l, s}\};\{ \theta^{0}_{ l'(v'), s}  
+ t (\theta^0_{l(v),s}  -\theta^{0}_{l'(v'), s})\}) \crcr
&&+ [\prod_{p\neq l} 
C_{j_{p}} (\{\theta^\ext_{p, s}\};\{ \theta^{0}_{ p'(v'), s} + t(\theta_{p(v),s} - \theta_{p'(v'),s} )\})] \crcr
&& T_{l} \triangleright T_l \triangleright C_{j_l}(\{\theta^\ext_{l s}\};\{ \theta^{0}_{ l'(v'), s}  
+ t (\theta^0_{l(v),s}  -\theta^{0}_{l'(v'), s})\})  \Big]
 \Big)  \,\crcr
&& 
\Big[ \prod_{\ell \neq l } 
C_{i_\ell}(\{\theta_{\ell(v), s}\};\{ \theta_{\ell(v'), s}\})\Big]
\prod_{v\in \cV}  
\delta(\theta_{v, s}-\theta_{v, s'}) \Big\} .
\eea
The following statement holds
\begin{lemma}
\label{lemfacto4}
The internal contribution of $\bar A_4(G^{(k)}_i)[\{ \theta^\ext_{l,s}\};0]$, namely
\bea
\int \Big[
[\prod_{l=2p+1} d\theta^0_{l,1}]
[\prod_{l=2p} d\theta^0_{l,2}d\theta^0_{l,3}d\theta^0_{l,4}]\Big] 
\Big[ \prod_{\ell \neq l } 
C_{i_\ell}(\{\theta_{\ell(v), s}\};\{ \theta_{\ell(v'), s}\})\Big] 
\prod_{v\in \cV}  
\delta(\theta_{v, s}-\theta_{v, s'})
\eea
does not depends on the set of variables $\{\theta^0_{l'(v'), s}\}$ 
used in the interpolation moves. Furthermore, the second
contribution identically vanishes:
\bea
\frac{d}{dt}\bar A_4(G^{(k)}_i)[\{ \theta^\ext_{l,s}\};0]=0\;.
\eea
\end{lemma}
\noindent{\bf Proof.} The first claim can  be proved using
translation invariance along the lines of the proof of Lemma \ref{lemfacto6}.
Indeed, the main point here is that, once again, one of  
the external position on external faces can be absorbed by 
successive changes of variables along a face. We simply 
choose to gauge away the interpolated positions belonging
to $\{\theta^0_{l', s}\}$. 

The second claim can be proved using the parity of functions. 
To this end, we start by writing the said contribution as
\bea
&&
\frac{d}{dt}\bar A_4(G^{(k)}_i)[\{ \theta^\ext_{l,s}\};0]=\crcr
&&
 \int \Big\{ [ \prod_{\ell} d\theta_{\ell, s}]\;\;
\Big( \sum_{l=1}^4 [\prod_{p\neq l} 
C_{j_{p}} (\{\theta^\ext_{p, s}\};\{ \theta^{0}_{ p'(v'), s}\})] \crcr
&&\sum_{k=0}^2\Big[ 
\delta_{l,2k+1} (\theta^0_{l,1}  -\theta^{0}_{l', 1}+r)
\partial_{\theta;1} 
+
\delta_{l,2k} \sum_{s=2}^4 
(\theta^0_{l,s}  -\theta^{0}_{l', s}+r)
\partial_{\theta;s} \Big] C_{j_{l}} (\{\theta^\ext_{l, s}\};\{ \theta^{0}_{ l'(v'), s}  \})\,\Big)\crcr
&&
\Big[ \prod_{\ell \neq l } 
C_{i_\ell}(\{\theta_{\ell(v), s}\};\{ \theta_{\ell(v'), s}\})\Big]\prod_{v\in \cV}  
\delta(\theta_{v, s}-\theta_{v, s'}) \Big\} .
\eea
One notices that the internal contribution 
\bea
\int
[ \prod_{\ell} d\theta_{\ell, s}] (\theta^0_{l,s}  -\theta^{0}_{l', s}+r)
\Big[ \prod_{\ell \neq l } 
C_{i_\ell}(\{\theta_{\ell(v), s}\};\{ \theta_{\ell(v'), s}\})\Big]\prod_{v\in \cV}  
\delta(\theta_{v, s}-\theta_{v, s'})
\eea
does not depend on $\{\theta^0_{l',s}\}$ by translation invariance and factors from the external data. We make two successive change of variables such that, $\hat\theta^0_{l,s}= \theta^0_{l,s} +r$, in all corresponding sectors of the theory, and then,  for all lines $\ell$, 
$ (\theta_{\ell,s}  -\theta^{0}_{l', s}) = \tilde \theta^0_{\ell,s}$, and
the internal part becomes 
\bea
&&
\int_{-\pi}^{\pi} d\tilde \theta^0_{l,s} \; \tilde \theta^0_{l,s} 
C_{i_l}(\{ \tilde \theta^0_{l,s} ;\tilde\theta_{l(v), s}\};\{  \tilde \theta_{l(v'), s}\}) \crcr
&&
\int [ \prod_{\ell \neq l} d\theta_{\ell, s}] 
\Big[ \prod_{\ell \neq l } 
C_{i_\ell}(\{ \tilde \theta_{\ell(v), s}\};\{  \tilde \theta_{\ell(v'), s}\})\Big]\prod_{v\in \cV}  
\delta(\tilde\theta_{v, s}-\tilde\theta_{v, s'})\;.
\eea
The result of this integral is vanishing due to the parity
of all propagators (see \eqref{propinit}) and delta functions while
  $\tilde \theta^0_{l,s} $ is clearly odd.

\qed

\begin{lemma}
\label{lemrem4}
The remainder $R_4$ of the amplitude interpolation 
can be bounded by
\beq
|R_4| \leq K M^{-2( i(G^{(k)}_i) - e(G^{(k)}_i) )} M^{\omega(G^{(k)}_i)}  \;,\qquad
e(G^{(k)}_i)  = \sup_{l \, \text{external to}\; G^{(k)}_i } j_l \;,\qquad
 i(G^{(k)}_i) = \inf_{l \in G^{(k)}_i} i_l\;,
\eeq 
for some constant $K$.
\end{lemma}
\noindent{\bf Proof.} The proof starts by removing
all $r$ in the same manner as performed in the proof of Lemma 
\ref{lemrem6} using the periodicity of all kernels. 
Then, expanding the derivative in the propagators
of the $T^2$ form,
we can bound the second order products as
$|(\theta^0_{l_1,s_1} - \theta^0_{l'_1,s_1})(\theta^0_{l_2,s_2} - \theta^0_{l'_2,s_2})| \leq M^{-2i(G^{(k)}_i)}$ whereas each 
derivative $\partial_{\theta;s}C_{j_l}$ by (\ref{boundder}) yields a factor $M^{e(G^{(k)}_i)}$
(second order derivative will contribute twice, and so forth).
We collect these improvements and write, using internal decay
to remove the differences $t(\theta^0_{l,s} - \theta^0_{l',s})$
and dropping the integral in $t$, 
\bea
|R_4| &\leq& K M^{2e(G^{(k)}_i)} M^{-2i(G^{(k)}_i)}
 \int [ \prod_{\ell} d\theta_{\ell, s}] \; \Big\{ \crcr
&&\Big(
\sum_{l=1}^6 
[\prod_{q\neq l} 
 M^{2j_q} 
e^{-\delta M^{j_q} \sum_s |\theta^\ext_{q, s}
- \theta^{0}_{ q'(v'), s}|}]
M^{2e(G^{(k)}_i)} 
e^{-\delta M^{j_l} \sum_s |\theta^\ext_{l, s}
- \theta^{0}_{ l'(v'), s}|} \Big)
\,\crcr
&& 
\Big[ \prod_{\ell \neq l } 
C_{i_\ell}(\{\theta_{\ell(v), s}\};\{ \theta_{\ell(v'), s}\})\Big]
\prod_{v\in \cV}  
\delta(\theta_{v, s}-\theta_{v, s'}) \Big\} .
\eea
\qed

At this stage, we have proved that the local part of 
amplitude is linear and of the form of the initial vertex $V_{4;1}$. 
For the second kind of vertex appearing $V_{4;2}$ and permutations, the same analysis also applies.

\subsection{Renormalization of the two-point function}
\label{subsect:twopt}

We perform now the interpolation moves for external
legs of the two-point function of a subgraph $G^{(k)}_i$
defined by the one of the five lines of Table 1. 
Here, we will be dealing with a graph with boundary 
of the form of a mass type vertex of the kind $V_2$.

Let $\bar A_2(G^{(k)}_i)$  denote the amplitude associated
with $G^{(k)}_i$ equipped with external propagators
with external positions variables $\theta^{\ext}_{l,s}$,  $l=1,2,$ and $s=1,2,3,4,$ and scale $j_l$. The couples $(\theta^0_{l,s},\theta^\ext_{l,s})$ keep their earlier relationship and sense.

We use the formula (\ref{interpol}) in order to rewrite
$\theta^{0}_{2,s}$ according to the particular pattern 
of the $\varphi^2$ vertices:
 the couples $(\theta^0_{1,s},\theta^{0}_{2, s})$ 
are connected with respect to the strand indices $s=1,2,3,4$.

The amplitude $\bar A_2(G^{(k)}_i)[\{ \theta^\ext_{l,s}\}]$ of the subgraph $G^{(k)}_i$ with external propagators (with above characteristics)
is re-expressed using the modified amplitude $\bar A_2(G^{(k)}_i)[\{ \theta^\ext_{l,s}\};t]$ as
\bea
\bar A_2(G^{(k)}_i)[\{ \theta^\ext_{l,s}\};t=1] &=&
 \bar A_2(G^{(k)}_i)[\{ \theta^\ext_{l,s}\};t=0] 
+ \frac{d}{dt}\bar A_2(G^{(k)}_i)[\{ \theta^\ext_{l,s}\};t=0] \\
& +& \frac12\frac{d^2}{dt^2} \bar A_2(G^{(k)}_i)[\{ \theta^\ext_{l,s}\};t=0]  
+\frac12 \int_0^1 (1-t)^2\frac{d^3}{dt^3} \bar A_2(G^{(k)}_i)[\{ \theta^\ext_{l,s}\};t] \;,
\nonumber
\eea 
where we define
\bea
\bar A_2(G^{(k)}_i)[\{ \theta^\ext_{l,s}\};t] &= &
\int [ \prod_{\ell} d\theta_{\ell, s}] \; \Big\{
C_{j_1}(\{\theta^\ext_{1 s}\};\{ \theta^{0}_{ 1, s} \})
 C_{j_2}(\{\theta^\ext_{2,s}\};\{ \theta^{0}_{1, s}  
+ t (\theta^0_{2,s}  -\theta^{0}_{1, s})\}) \,\crcr
&& 
\Big[ \prod_{\ell \neq l } 
C_{i_\ell}(\{\theta_{\ell(v), s}\};\{ \theta_{\ell(v'), s}\})\Big]
\prod_{v\in \cV}  
\delta(\theta_{v, s}-\theta_{v, s'})\Big\} . 
\eea
The different quantities involved in the expansion can be 
studied. 
The first contribution is of the form
\bea
&& \bar A_2(G^{(k)}_i)[\{ \theta^\ext_{l,s}\};0]=   \int \Big\{ [ \prod_{\ell \neq l} d\theta_{\ell, s}]\,
\Big[ 
\prod_{l=1}^2 C_{j_l}(\{\theta^\ext_{l, s}\};\{ \theta^{0}_{ l'(v'), s}\})\Big] 
\crcr
&&\int \Big[ \prod_{s} d\theta^0_{2,s}\Big] 
\Big[ \prod_{\ell \neq l } 
C_{i_\ell}(\{\theta_{\ell(v), s}\};\{ \theta_{\ell(v'), s}\})\Big] \prod_{v\in \cV}  
\delta(\theta_{v, s}-\theta_{v, s'}) \Big\}\crcr
&&
\prod_{l=1}^2 C_{j_l}(\{\theta^\ext_{l, s}\};\{ \theta^{0}_{ l'(v'), s}\}) = \crcr
&&
 C_{j_1}(\{\theta^\ext_{1,s}\};\{\theta^{0}_{1, 1}, \theta^{0}_{1,2} ,\theta^{0}_{1,3}, \theta^{0}_{1,4}\})
C_{j_2}(\{\theta^\ext_{2,s}\};\{\theta^{0}_{1, 1}, \theta^{0}_{1,2} ,\theta^{0}_{1, 3}, \theta^{0}_{1,4}\})   \;.
\label{at02}
\eea
The $\alpha$-th derivative terms, $\alpha=1,2$, are given by
\bea
 \frac{d^\alpha}{dt^\alpha}\bar A_2(G^{(k)}_i)[\{ \theta^\ext_{l,s}\};t=0] 
&=&
  \int \Big\{ [ \prod_{\ell} d\theta_{\ell, s}]\;\;
\Big(  
C_{j_{1}} (\{\theta^\ext_{1, s}\};\{ \theta^{0}_{ 1, s}\})] 
\, T^\alpha \,\triangleright  C_{j_{2}} (\{\theta^\ext_{2, s}\};\{ \theta^{0}_{ 1, s}  \})\Big)\crcr
&&
\Big[ \prod_{\ell \neq l } 
C_{i_\ell}(\{\theta_{\ell(v), s}\};\{ \theta_{\ell(v'), s}\})\Big]\prod_{v\in \cV}  
\delta(\theta_{v, s}-\theta_{v, s'}) \Big\} ,
\label{rmaind21}
\eea
where $T^\alpha$ stands for the operator, using previous notations,
\bea
T^\alpha :=  \sum_{s_\alpha} 
\prod_{\alpha} (\theta^0_{2,s_\alpha}  -\theta^{0}_{1,s_\alpha} +r)
\prod_{\alpha} \partial_{\theta;s_\alpha} \;,  \qquad 
\alpha= 1,2,3, \;\quad  s_{\alpha}=1,2,3,4\;.
\eea
 Last, the remainder can be written as
\bea
R_2 &=& \frac12 \int_{0}^1 dt \;(1-t)^2 \int [ \prod_{\ell} d\theta_{\ell, s}] \; \Big\{ \crcr
&&\Big(
C_{j_{1}} (\{\theta^\ext_{1, s}\};\{ \theta^{0}_{1, s}\})]\;
 \,T^3 \,\triangleright\; C_{j_2}(\{\theta^\ext_{2,s}\};\{ \theta^{0}_{1, s}  
+ t (\theta^0_{2,s}  -\theta^{0}_{1, s})\}) \Big)   \,\crcr
&& 
\Big[ \prod_{\ell \neq l } 
C_{i_\ell}(\{\theta_{\ell(v), s}\};\{ \theta_{\ell(v'), s}\})\Big]
\prod_{v\in \cV}  
\delta(\theta_{v, s}-\theta_{v, s'}) \Big\} \;.
\eea
The main properties of the different parts of the expansion 
are summarized in the following propositions. 
\begin{lemma}
\label{lemfacto2}
The internal contribution of $\bar A_2(G^{(k)}_i)[\{ \theta^\ext_{l,s}\};0]$, namely
\beq
\int \Big[
[\prod_{s} d\theta^0_{2,s} ] 
\Big[ \prod_{\ell \neq l } 
C_{i_\ell}(\{\theta_{\ell(v), s}\};\{ \theta_{\ell(v'), s}\})\Big] 
\prod_{v\in \cV}  
\delta(\theta_{v, s}-\theta_{v, s'})
\eeq
does not depends on the set of variables $\{\theta^0_{1, s}\}$ 
used in the interpolation moves. Furthermore, we have
\beq
\frac{d}{dt}\bar A_2(G^{(k)}_i)[\{ \theta^\ext_{l,s}\};0]=0\;,
\eeq
and the third term reduces to
\beq
\frac{d^{2}}{dt^{2}}\bar A_2(G^{(k)}_i)[\{ \theta^\ext_{l,s}\};0]
= \log M 
\int [ \prod_{s} d\theta_{1, s}]\;\;
C_{j_{1}} (\{\theta^\ext_{1, u}\};\{ \theta^{0}_{ 1, u}\}) 
\sum_{s=1}^4 \Delta_{s}^2 C_{j_{2}} (\{\theta^\ext_{2, v}\};\{ \theta^{0}_{ 1, v}  \})\;,
\eeq
where $ \Delta_{s}$ is a Laplace operator on $U(1)$ acting
on the strand $s$.
\end{lemma}
\noindent{\bf Proof.} The first claim is a consequence
of translation invariance and provides a mass renormalization.
The second claim can be proved using the parity of propagators
along the lines of the proof of Lemma \ref{lemfacto4}.
Let us focus on the last claim and write (using more
symbols in order to differentiate strand indices)
\bea
 && \frac{d^2}{dt^2}\bar A_2(G^{(k)}_i)[\{ \theta^\ext_{l,s}\};t=0] 
=\crcr
&& \sum_{s,s'=1}^4 \int \Big\{ [ \prod_{s} d\theta_{1, s}]\;\;
\Big(  
C_{j_{1}} (\{\theta^\ext_{1, u}\};\{ \theta^{0}_{ 1, u}\})\, 
\partial_{\theta;s} \partial_{\theta;s'}  \,C_{j_{2}} (\{\theta^\ext_{2, w}\};\{ \theta^{0}_{ 1, w}  \})\Big)\crcr
&&\int  [ \prod_{\ell \neq l} d\theta_{\ell, s}]
\Big[  (\theta^0_{2,s}  -\theta^{0}_{1, s}+r)(\theta^0_{2,s'}  -\theta^{0}_{1, s'}+r) \prod_{\ell \neq l } 
C_{i_\ell}(\{\theta_{\ell(v), q}\};\{ \theta_{\ell(v'), q}\})\Big]\crcr
&&\prod_{v\in \cV}  
\delta(\theta_{v, s}-\theta_{v, s'}) \Big\} \;.
\eea
We first implement a change of variable which removes all 
$r$. Next,  we use translation invariance in order to remove 
from the internal part the dependence in $\theta^0_{1,s}$. 
One gets
\beq
\int  [ \prod_{\ell \neq 1} d\tilde\theta_{\ell, s}]
\Big[  \tilde\theta^0_{2,s}  \tilde\theta^0_{2,s'}  \prod_{\ell \neq l } 
C_{i_\ell}(\{\tilde\theta_{\ell(v), s}\};\{\tilde \theta_{\ell(v'), s}\})\Big]\prod_{v\in \cV}  
\delta(\tilde\theta_{v, s}-\tilde\theta_{v, s'}) \;.
\eeq
When $s\neq s'$, the above integral vanishes because of the parity 
of the integrand function, recall that $\tilde\theta^0_{2,s}$ and $\tilde\theta^0_{2,s'} $ belong to $[-\pi,\pi)$. Only remains the terms at $s=s'$ which are
\bea
&&
 \frac{d^2}{dt^2}\bar A_2(G^{(k)}_i)[\{ \theta^\ext_{l,s}\};t=0] 
= \sum_{s=1}^4 
\int \Big\{ [ \prod_{s} d\theta_{1, s}]\;\;
\Big(  
C_{j_{1}} (\{\theta^\ext_{1, u}\};\{ \theta^{0}_{ 1, u}\}) 
\partial_{\theta;s}^2 C_{j_{2}} (\{\theta^\ext_{2, v}\};\{ \theta^{0}_{ 1, v}  \})\Big)\crcr
&& \times \int 
[ \prod_{\ell \neq 1} d\tilde\theta_{\ell, s}] \Big[ (\tilde\theta^0_{2,s})^2\prod_{\ell \neq 1 } 
C_{i_\ell}(\{\tilde\theta_{\ell(v), u}\};\{\tilde \theta_{\ell(v'), u}\})\Big]\prod_{v\in \cV}  
\delta(\tilde\theta_{v, s}-\tilde\theta_{v, s'}) \Big\} \;.
\eea
In fact, the internal part does not depend on the strand index 
$s$ because, from the beginning, all $s=1,2,3,4$ are treated
in a symmetric manner. 
Hence all integrations as $\int d\tilde\theta^0_{2,s}$, 
for any $s$, should produce the same result. 
Moreover, the $\tilde\theta^0_{2,s}$ factors are of order $M^{-2i_2}$. This contribution cancels the internal
quadratic divergence (same as for the mass local part) thus 
yielding a logarithmic divergence for a wave function renormalization.

\qed

\begin{lemma}
\label{lemrem2}
The remainder $R_2$ of the amplitude interpolation 
can be bounded by
\beq
|R_2| \leq K M^{-3( i(G^{(k)}_i) - e(G^{(k)}_i) )} M^{\omega(G^{(k)}_i)} \;,\qquad
e(G^{(k)}_i)  = \sup_{l \, \text{external to}\; G^{(k)}_i } j_l \;,\qquad
 i(G^{(k)}_i) = \inf_{l \in G^{(k)}_i} i_l\;,
\eeq 
for some constant $K$.
\end{lemma}
\noindent{\bf Proof.} 
As in the earlier setting, in the convenient variables,
the $T^3$ operator applied on the propagators
yields a prefactor of the form $|(\theta^0_{2,s_k} - \theta^0_{1,s_k})^3|$
which can be bounded by $ M^{-3i(G^{(k)}_i)}$ whereas each 
derivative $\partial_{\theta;s}C_{j_l}$  yields a 
good factor $M^{e(G^{(k)}_i)}$ according (\ref{boundder}).
We  infer the bound (removing the differences $t(\theta^0_{l,s} - \theta^0_{l',s})$ due to strong internal decay) 
\bea
|R_2| &\leq& K M^{-3(i(G^{(k)}_i)-e(G^{(k)}_i))}
 \int [ \prod_{\ell} d\theta_{\ell, s}] \; \Big\{ \crcr
&&\Big(
\sum_{l=1}^6
[\prod_{q\neq l} 
 M^{2j_q} 
e^{-\delta M^{j_q} \sum_s |\theta^\ext_{q, s}
- \theta^{0}_{ q'(v'), s}|}]
M^{2e(G^{(k)}_i)} 
e^{-\delta M^{j_l} \sum_s |\theta^\ext_{l, s}
- \theta^{0}_{ l'(v'), s}|} \Big)
\,\crcr
&& 
\Big[ \prod_{\ell \neq l } 
C_{i_\ell}(\{\theta_{\ell(v), s}\};\{ \theta_{\ell(v'), s}\})\Big]
\prod_{v\in \cV}  
\delta(\theta_{v, s}-\theta_{v, s'}) \Big\} .
\eea
\qed

In summary, the expansion of the two-point function
gives a local contribution to the mass which is quadratically
divergent, a wave function renormalization which is logarithmically
divergent and a remainder which will allow to sum on the 
momentum assignments. 

The fact that the theory is renormalizable at all order of perturbations then follows from the standard techniques of summation on
momentum assignments developed in \cite{Rivasseau:1991ub}. Remark that the theory is well-prepared by the multiscale expansion to  
be written in terms of an infinite set of effective couplings which follow the renormalization group trajectory. It is almost a pity to
add the counterterms corresponding to non quasi-local subgraphs to re-express the theory in terms of the standard renormalized couplings.
This  does not of course introduce any divergence and the coefficients of that renormalized power series
can be proved term by term finite. However the renormalized series is in fact much less natural 
than the effective one, and  is plagued by  large undesirable 
contributions called renormalons. These phenomena, analyzed at length in 
\cite{Rivasseau:1991ub}, will not be further discussed here.

\section{Conclusion}

The tensor model presented here is not claimed to be the right final model for quantum gravity but hopefully a first step in that direction.

It also completes nicely the progressive discovery of new forms of renormalization group
with different types of divergent graphs. There seems to be a natural hierarchy of these forms.
In ordinary just renormalizable models such as the local $\phi^4_4$ theory or Yang-Mills theory, the divergence degree
 is simply a function of the number of external legs.
In the condensed matter theory of interacting electrons in any dimension, the renormalization
group is already very different. It is governed by the approach to the Fermi surface, which is a codimension 2 singularity in space-time. 
In this particular instance, the important flow is that of
the Cooper pair four-point coupling constant in the $s=0$ channel. Only a very simple category
of four-point graphs contribute to that flow, namely those which are chains of bubbles. They are also the ones
leading in the $1/N$ expansion for vector models and form a geometric series. 
The BCS transition can be analyzed accordingly. $N$ can be interpreted as the number of sectors or quasi particles
around the Fermi surface \cite{Feldman:1993ck}.  We propose to consider the renormalization group for such models as vector-like.

In the Grosse-Wulkenhaar model \cite{Grosse:2004yu,Rivasseau:2005bh}
as well as the $\phi^6_\star$-theory as developed in \cite{zhituo}, the non-commutativity of the underlying space-time
translates into a matrix representation of the theory with a perturbative expansion indexed by ribbon graphs.
The divergent graphs are the planar graphs with all external legs incident on a single face. They are the ones leading the $1/N$ matrix expansion. 
Obviously the renormalization group for such models should be called matrix-like.

The models of this paper pioneers a new category of renormalization group,  based on tensor fields of rank higher than 2.
The key divergent graphs are the melonic ones. Their renormalization group should be called of tensor-type.
This issue and the systematic study of such models including their 
renormalization group flows, symmetries (either in the spirit of \cite{BenGeloun:2011cz,BenGeloun:2011xu}
or in that of \cite{Baratin:2011tg}), the issue of their constructive stability and possible phase transitions is left to future studies
 (the corresponding program and its relationship to other approaches to quantum gravity is further discussed in \cite{Rivasseau:2011hm}).

As a final remark, we conjecture that if we restrict the couplings of the model studied above to the precise
values given by the integration of the standard colored theory, that is if we link $\lambda_{6;1}$
 $\lambda_{6;2}$ and $\lambda_{4;1}$ as they should be when integrating the $D$ fields of a colored tensor 
theory with single coupling $\lambda$, we should obtain an even more interesting just renormalizable theory
with a single coupling, i.e. the corresponding manifold should be stable under the renormalization group flow.

\section*{ Appendix}
\label{app}

\appendix

\renewcommand{\theequation}{\Alph{section}.\arabic{equation}}
\setcounter{equation}{0}

\section{Propagator bounds}
\label{app:prop}

We consider the propagator in the slice $i$ as
\bea
C_i(\{\theta_{s}\}; \{\theta'_s\}) &=& k
\int_{M^{-2(i+1)}}^{M^{-2i}} \frac{e^{-m^2\alpha  }}{\alpha ^2}
e^{ - \frac{1}{4\alpha }\sum_s  [\theta_s -  \theta'_{s}]^2 } \;
 T(\alpha;\{\theta_{s}\}; \{\theta'_s\})  \; d\alpha\;,\crcr
T(\alpha;\{\theta_{s}\}; \{\theta'_s\})
&=&\prod_{s=1}^4 \left\{ 
1+ 2\sum_{n=1}^\infty e^{-\frac{\pi^2n^2}{\alpha}} \cosh\Big[  \frac{n\pi}{\alpha}[\theta_s -  \theta'_{s}]\Big]
\right\} \;.
\label{propoi}
\eea
Since only positive terms are involved in each series in the product $T$,
we can find an integral bounding the series as 
\beq
\sum_{n=1}^\infty e^{-\frac{\pi^2n^2}{\alpha}} \cosh\Big[  \frac{n\pi}{\alpha}\Theta\Big]
 \leq \int_{1}^\infty e^{-\frac{\pi^2x^2}{\alpha}} \cosh\Big[  \frac{\pi \Theta}{\alpha}x \Big] dx\;,
\label{integ}
\eeq
where $|\Theta|< 2\pi$. The latter integral can be recast in terms of  Gaussian error  functions: 
\bea
&&
\int_{1}^\infty 
e^{-\frac{\pi^2x^2}{\alpha}} \cosh\Big[  \frac{\pi \Theta}{\alpha} \Big]dx = 
\frac{\sqrt{\alpha }
   e^{\frac{ \Theta^2}{4 \alpha }} }{4\sqrt{\pi }}\left(\text{erfc}
\left(\frac{2 \pi - \Theta}{2 \sqrt{\alpha
   }}\right)+\text{erfc}\left(
   \frac{ 2 \pi +\Theta}{2
   \sqrt{\alpha
   }}\right)\right) ,
\label{erfc}\\
\crcr
&&
\text{erfc}(z)  = 1 - \frac{2}{\sqrt \pi}\int_{0}^z e^{-t^2} dt  \;, \quad z \in \C\;; \quad     \text{erfc}(x) \leq \frac{2}{\sqrt{\pi}} \frac{e^{-x^2}}{x+ \sqrt{x^2 + \frac{4}{\pi}}} \leq e^{-x^2}\;,\quad x >0\;,\nonumber
 \eea
therefore, given $ -2\pi <\Theta <2\pi$,
\bea
&&
   e^{\frac{ \Theta^2}{4 \alpha }}\left(\text{erfc}
\left(\frac{2 \pi - \Theta}{2 \sqrt{\alpha
   }}\right)+\text{erfc}\left(
   \frac{ 2 \pi +\Theta}{2
   \sqrt{\alpha
   }}\right)\right) \leq  
 e^{\frac{ \Theta^2}{4 \alpha }} \left[
e^{-\left(\frac{2 \pi - \Theta}{2 \sqrt{\alpha }}\right)^2}+
e^{-\left(\frac{ 2 \pi +\Theta}{2 \sqrt{\alpha} } \right)^2 } \right] \crcr
&& 
\leq  e^{\frac{- \pi^2 }{ \alpha}}
\left[
e^{\frac{ \pi\Theta}{2 \alpha}}+
e^{\frac{-\pi\Theta}{2\alpha }  } \right] \leq 2\;.
\label{erc}
\eea
Then, combining \eqref{erc}, \eqref{erfc} and \eqref{integ}, 
we find a bound for $T$ \eqref{propoi} as
\beq
T(\alpha;\{\theta_{s}\}; \{\theta'_s\})   \leq  \prod_{s=1}^4 \left\{ 
1+ \frac{\sqrt{\alpha }}{\sqrt{\pi }}
\right\}\;,
\eeq
hence the following bound is achieved
\beq
C_i(\{\theta_{s}\}; \{\theta'_s\}) \leq  k
\int_{M^{-2(i+1)}}^{M^{-2i}} \frac{e^{-m^2\alpha  }}{\alpha ^2}
e^{ - \frac{1}{4\alpha }\sum_s  [\theta_s -  \theta'_{s}]^2 } 
  \prod_{s=1}^4 \left\{ 
1+ \frac{\sqrt{\alpha }}{\sqrt{\pi }} \right\} d \alpha\;.
\eeq
By expanding the product, it can be observed that the term 
with coefficient 1 is the dominant one. The sum of remaining terms,
including powers of $\sqrt\alpha$ in their numerator, 
can be bounded by a constant (mainly, the number of terms) times  the
leading term. Indeed, for instance, focusing on the subleading term of the form
\bea
 k
\int_{M^{-2(i+1)}}^{M^{-2i}} \frac{e^{-m^2\alpha  }}{\alpha ^2}
e^{ - \frac{1}{4\alpha }\sum_s  [\theta_s -  \theta'_{s}]^2 } 
   \frac{\sqrt{\alpha }}{\sqrt{\pi }}   d \alpha   &=&
k'
\int_{M^{-2(i+1)}}^{M^{-2i}} \frac{e^{-m^2\alpha  }}{\alpha ^{\frac{3}{2}}}
e^{ - \frac{1}{4\alpha }\sum_{s}  [\theta_s -  \theta'_{s}]^2 } 
  d \alpha \crcr
&\leq&  k' \int_{M^{-2(i+1)}}^{M^{-2i}} \frac{e^{-m^2\alpha  }}{\alpha ^2}
e^{ - \frac{1}{4\alpha }\sum_s  [\theta_s -  \theta'_{s}]^2 } \;.
\nonumber
\eea
Higher order terms involve $\alpha^{\frac{d}{2}}$, $d \geq 1$, in the numerator, hence they will be less divergent. 
This validates the bound $C_i$  \eqref{ci} for all $i \gg 1$. 

For the last slice, we have
\beq
C_0(\{\theta_{s}\}; \{\theta'_s\}) \leq  k
\int_{1}^{\infty} \frac{e^{-m^2\alpha  }}{\alpha ^2}
e^{ - \frac{1}{4\alpha }\sum_s  [\theta_s -  \theta'_{s}]^2 } 
  \prod_{s=1}^4 \left\{ 
1+ \frac{\sqrt{\alpha }}{\sqrt{\pi }} \right\} d \alpha\;. 
\eeq
This expression can be bounded, this time, by the term containing
the highest power of $\sqrt{\alpha}$:
\bea
C_0(\{\theta_{s}\}; \{\theta'_s\}) &\leq&  K'
\int_{1}^{\infty} \frac{e^{-m^2\alpha  }}{\alpha ^2}
e^{ - \frac{1}{4\alpha }\sum_s  [\theta_s -  \theta'_{s}]^2 } 
  \alpha^2 d \alpha \crcr
& \leq& K'
\sup_{\alpha \in [1,+\infty)}
\Big(e^{-m^2\alpha/2  } \alpha^2
e^{ - \frac{1}{4\alpha }\sum_s  [\theta_s -  \theta'_{s}]^2 } \Big)
\int_{1}^{\infty} \frac{e^{-m^2\alpha/2  }}{\alpha^2}
 d \alpha \crcr
&\leq& K'' 
\int_{1}^{\infty} \frac{e^{-m^2\alpha/2  }}{\alpha^2}
 d \alpha 
\eea
which validates \eqref{coin}.

\section{The Three Dimensional Case}

In three dimensions, there is also a just renormalizable similar model but with  propagator $(\sum_{s=1}^3 |p_s| +m)^{-1}$
and a single melonic ``pillow" interaction
\bea S_4 = \int_{h_j} \psi_{1,2,3} \,\bar\psi_{1',2,3}\,\psi_{1',2',3'} \,\bar\psi_{1,2',3'}\, 
+ \text{permutations }  \;  .
\label{fifi1}
\eea
As usual, we have to introduce a mass counterterm $V_2$ and a  $V'_2$ wave function $\sum_{s=1}^3 |p_s|$ counterterm. 

The scaling of the sliced propagator is now
\bea  C^i  \le K M^{2i} e^{-\delta M^i \sum_{s}\vert \theta_s - \theta'_s \vert}\;,
\eea
hence the power counting is $\omega_d  =  -L +F + V'_2 $. Following the analysis of Section 4 and 5, with same notations we have
$4 V_4  +2 (V_2  + V'_2 ) = 2L + N_{\ext}$ and
\beq
V_{\cexG} = 4 V_4  + 2 (V_2  + V_2')  \;, \quad
L_{\cexG} =  L + L_{\inter;\,\cexG} = \frac12 (4 V_{\cexG} - N_{\ext})\;,
\eeq
There are 3 jackets in $\cexG$. Each face of the graph $\cexG$ (open or closed) is shared by exactly 2 jackets so that
$\sum_{J} F_{J} = 2 F_{\cexG}$ and 
\beq
\sum_{J} ( F_{\inter;\,   \tJ; \, \cG} + F_{\inter;\,   \tJ; \, \cexG} + F_{\ext; \tJ}) = 
2 F_{\inter;\cG} + 2 F_{\inter;\;  \cexG}  + \sum_{J} F_{\ext; \tJ}\;.
\eeq
Each $\varphi^4$ vertex contains $4$ internal faces and each
$\varphi^2$ vertex contains $3$ internal faces
so that
$
F_{\inter;\; \cexG}  = 4 V_4+ 3(V_2 +  V_2')\;.
$
Hence
\beq
\sum_{ J}\left[ - V_{J} + L_{J}\right]  =
  3[ 4V_4  +2(V_2  +  V_2' ) ]- \frac32 N_{\ext}\;, 
\eeq
and 
\beq
F_{\inter;\cG}  =2 V_4  -  \frac34 N_{\ext} + 3 - \sum_{J} g_{\tJ}   -  \frac12  \sum_{J} F_{\ext; \tJ} \;.
\label{facint1}
\eeq
The boundary graph is a closed ribbon graph
living in dimension $D-1 = 2$, hence has a single jacket.
Since each external leg of the initial
graph $\cG$ has 3 strands and an external leg is made with 
two end-points belonging to two external legs, we have in this simpler case
\bea
L_{\bG} - V_{\bG} = \frac{1}{2}  N_{\ext}\;, \;
F_{\bG}  =2(C_{\bG}-1) -2g_{\bG} + 2 + \frac{1}{2}  N_{\ext}\;, \;
\sum_{J} F_{\ext; \,\tJ} = F_{\bG}\; ,
\eea
(with again $ g_{\bG} = \sum_{\rho}g_{{_{\bG} }_{\rho}}$) and finally we get the divergence degree
\beq
\omega_d(\cG)=
 -V_2  - \frac12  ( N_{\ext} -4)  - \sum_{J} g_{\tJ}  + g_{\bG}    -  (C_{\bG}-1)\;,
\eeq
appearing as a simpler analog of Theorem \ref{convtop} and formula \eqref{contopformula}. The complete proof of the renormalizability of this model following Section 5 and 6 has been addressed in a recent\footnote{  The said manuscript was published shortly after the publication of the first version of this paper.} work \cite{BenGeloun:2012pu}.
\qed 

\section*{Acknowledgements}
We thank R. Gurau for useful discussions at various stages 
of this work and for a critical reading of the manuscript.
The authors also thank the referee for his careful 
reading and interesting remarks improving Lemma 4 
and the presentation of this paper. 

Research at Perimeter Institute is supported by the Government of Canada through Industry
Canada and by the Province of Ontario through the Ministry of Research and Innovation.

\end{document}